\documentclass[10pt,a4paper]{article}
\usepackage{amsfonts}
\usepackage{amssymb}
\usepackage{amsmath}
\usepackage{textcomp}
\usepackage{bbm}
\usepackage{stmaryrd}
\usepackage{bm}

\usepackage{graphicx}
\usepackage{theorem}
\newcommand{\qed}{\hfill$\Box$}

\def \HHH{\mathcal H}

\def \E{\mathcal E}
\def \NN{\mathbbm N}
\def \NNZ{\mathbbm N_0}
\def \ZZ{\mathbbm Z}
\def \ZZ{\mathbbm Z}

\def \RR{\mathbbm R}

\def \DD{\mathcal D}

\def \A{\mathcal A}

\def \EE{\mathbbm E}

\newcommand{\proof}{\noindent\emph{Proof: }}

{\theorembodyfont{\rmfamily} }
{\theorembodyfont{\rmfamily} }
\newtheorem{theorem}{Theorem}[section]

\newtheorem{lemma}{Lemma}[section]

\newtheorem{cor}{Corollary}[section]
\newcommand{\ra}{\rightarrow}

\newcommand{\D}{\Delta}
\newcommand{\g}{\gamma}

\newcommand{\subs}{\subseteq}

\newcommand{\dv}{{\rm{div}}}

\newcommand{\wt}{\widetilde}
\newcommand{\ee}{\varepsilon}

\newcommand{\phih}{\varphi}
\newcommand{\mx}{\vee}
\newcommand{\mn}{\wedge}
\newcommand{\x}{\times}

\newcommand{\s}{\sigma}
\newcommand{\ls}{\langle}
\newcommand{\rs}{\rangle}

\newcommand{\sm}{\setminus}
\newcommand{\bs}{\boldsymbol}
 \newcommand{\Z}{\mathcal Z} 
 \newcommand{\bbar}{\overline}  \newcommand{\fr}{\frac{1}}
\newcommand{\PP}{\mathbbm{P}}  \newcommand{\T}{\mathbbm{T}} \newcommand{\MM}{\mathbbm{M}}
 \newcommand{\1}{\mathbbm{1}}  
   \newcommand{\y}{\upsilon}
\newcommand{\pd}{\partial}   
 \newcommand{\R}{\mathcal{R}} \makeatletter \newcommand{\esssup}{\mathop{\operator@font ess\,sup}}
\makeatother \makeatletter \newcommand{\argmin}{\mathop{\operator@font argmin}}

\makeatother
\title{Hydrodynamic limit of condensing two-species zero range processes with sub-critical initial profiles} \author{Nicolas Dirr,
  Marios G.~Stamatakis and Johannes Zimmer}
\begin{document}
	
\maketitle
\begin{abstract}
  Two-species condensing zero range processes (ZRPs) are interacting particle systems with two species of particles and zero range
  interaction exhibiting phase separation outside a domain of sub-critical densities. We prove the hydrodynamic limit of nearest
  neighbour mean zero two-species condensing zero range processes with bounded local jump rate for sub-critical initial profiles,
  i.e., for initial profiles whose image is contained in the region of sub-critical densities. The proof is based on H.~T.~Yau's
  relative entropy method, which relies on the existence of sufficiently regular solutions to the hydrodynamic equation. In the
  particular case of the species-blind ZRP, we prove that the solutions of the hydrodynamic equation exist globally in time and
  thus the hydrodynamic limit is valid for all times.
\end{abstract}

\section{Introduction} 
\label{sec:Introduction}

In this article, we derive the hydrodynamic limit of a system of two interacting particle systems, specifically two-species zero
range processes (ZRPs). The motivation for this study is that hydrodynamic limits provide effective descriptions of large scale
interacting particle systems. There is a now good understanding of this limit passage for a range of particle processes leading to
one hydrodynamic limit equation. In particular, Kipnis and Landim~\cite{Kipnis1999a} establish the hydrodynamic behaviour for the
one-species zero range process, using the entropy method of Guo, Papanicolaou and Varadhan~\cite{Guo1988a}. For systems, however,
this limit passage is less well studied, and several tools available for single equations are no longer available, as explained in
more detail below. In particular, many systems where a hydrodynamic passage would be of interest both in its own right and as a
tool to understand the limiting system of partial differential equations (PDEs) are currently inaccessible to the methods
available; the full Patlak Keller-Segel system~\cite{Hillen2009a} modelling the evolution of cells or bacteria guided by the
concentration of a chemical substance is an example.  Yet, there are several recent studies focusing on different models of
interacting particle systems. One avenue is to derive equations which incorporate aspects of underlying models, be it by
considering the motion of cells only in a stationary, but random, environment mimicking the chemical~\cite{Grosskinsky2015a}, or
by an equation with a singular potential related to a Green's function describing the solution of a second
equation~\cite{Godinho2015a}. The hydrodynamic limit system of an active exclusion process modelling active matter has been
recently derived using a two-block estimate and non-gradient estimates~\cite{Clement2016a}.

Another approach is to study systems related to underlying zero-range processes (ZRPs) of several species to obtain a limiting
system, and this is the approach we pursue here. The focus on ZRPs can be motivated by their nature as a toy model of an
interacting particle system. We consider a system of two zero-range processes but the extension to $n$ types is
straightforward. Each ZRP is a process on a lattice where particles jump from one site to another according to a jump rate
function depending on the number of the two species of particles on this site only (hence the name zero range).

The hydrodynamic limit in the Eulerian scaling $t\mapsto tN$ of asymmetric many-species ZRPs with product and translation
invariant equilibrium states has been studied in~\cite{Grosskinsky2003b}. The hydrodynamic limit in the parabolic scaling
$t\mapsto tN^2$ for a class of processes not satisfying the assumptions of~\cite{Grosskinsky2003b} has also recently been
studied~\cite{Tsunoda2016a}; there one type of particles performs a random walk and influences the other type, which is a process
of ZRP type. In general, establishing hydrodynamic limits for systems of equations rigorously is a hard problem, with few known
results so far. To name a few, the hydrodynamic limit of a two-species simple-exclusion process was first studied
in~\cite{Quastel1992a}, the Leroux system has been derived as a hydrodynamic limit in~\cite{Toth2004}, and hyperbolic systems have
also been studied in~\cite{Toth2005}.

Here we consider a system of two ZRPs. We show that the hydrodynamic equation is a quasilinear parabolic system of the form
\begin{equation}
  \label{NonLinDiffSystem}
  \pd_t\bs{\rho}=\Delta\bs{\Phi}(\bs{\rho}),\quad\bs{\rho}=(\rho_1,\rho_2)\colon[0,T)\x\T^d\to\RR_+^2,
\end{equation} 
where $\pd_t\bs{\rho}:=(\pd_t\rho_1,\pd_t\rho_2)$, $\Delta\bs{\Phi}(\bs{\rho}):=(\Delta\Phi_1(\bs{\rho}),\Delta\Phi_2(\bs{\rho}))$
and $\bs{\Phi}$ is the mean jump rate of the ZRP at a site $x\in\T_N^d$ under the product and translation invariant equilibrium
state of background density $\bs{\rho}\in\RR_+^2$. Two-species ZRPs, and the phase transition they exhibit, were first studied
in~\cite{Evans2003a}. In condensing ZRPs, the set of admissible background densities $\bs{\rho}$ is a strict subset of
$\RR_+^2$. We call such densities \emph{sub-critical}.

One challenge of ZRPs is that they can exhibit condensation phenomena, where particles congregate at the same
site~\cite{Grosskinsky2003a,Evans2000a,Evans2005a}. Even for one-species systems, the hydrodynamic limit of ZRPs experiencing
condensation is presently unknown. We consider parameter regimes of two-species systems where condensation can occur, but restrict
to \emph{sub-critical initial profiles}, i.e., initial data that take values in the set of sub-critical densities. For one-species
ZRPs, the analogous result has been established recently~\cite{Stamatakis2015a} and we extend this argument to the two-species
case. Specifically, we apply the Relative Entropy method of H.~T.~Yau~\cite{Yau1991a}, which requires only the one-block estimate
proved in Theorem~\ref{OBE}, not the full replacement lemma~\cite[Lemma 5.1.10]{Kipnis1999a}. Thus it does not require the
equilibrium states of the ZRP to have full exponential moments, a property not satisfied by condensing ZRPs. This extension
of~\cite{Stamatakis2015a} is non-trivial, for two reasons. The first difficulty is that the relative entropy method requires the
existence of $C^{2+\theta}$ solutions to~\eqref{NonLinDiffSystem} for some $\theta\in(0,1]$, and that the solution remains in the
sub-critical region. By a result of Amann~\cite{Amann1986a} $C^{1,2+\theta}$ solutions exist locally in time, i.e., for small time
intervals when starting from $C^{2+\theta}$ initial data. By continuity, there is a non-trivial time interval such that a
solution, when starting from the sub-critical region, remains in the sub-critical region. So the general result on the
hydrodynamic limit is local in time, being valid for the largest time interval on which we have $C^{2+\theta}$ solutions in the
sub-critical region. This result shows that as long as a classical $C^{1,2+\theta}$ solution to the hydrodynamic equation exists,
condensation does not occur. The second difficulty to extend the results for one species~\cite{Kipnis1999a,Stamatakis2015a} is
that the phase space $\RR_+^2$ is now more complicated, and the one-dimensional arguments used
in~\cite{Kipnis1999a,Stamatakis2015a} do not extend directly. In particular, a novel argument is required to extend~\cite[Lemma
6.1.10]{Kipnis1999a}; see Lemma~\ref{LastBound} and its proof. Specifically, we employ a characterisation of the domain of a
convex function via the recession function of its Legendre transform. This characterisation of the domain of convex functions is
of interest in its own right in the context of two-species ZRPs. For example, it immediately yields a parametrisation of the
boundary of the domain of the partition function via the recession function of the thermodynamic entropy.

Intuitively, condensation means on the level of the governing hydrodynamic limit PDE the formation of singularities where the mass
concentrates. For scalar equations, the formation of such singularities can be ruled out by a maximum principle. For systems,
however, in general maximum principles do not hold. In this article, we mainly rely on an existence theory for local
$C^{1,2+\theta}$ classical solutions established by Amann and focus on proving the local in time hydrodynamic limit. However, for
a particular example, the so-called \emph{species-blind process}, we are able to establish a maximum principle and
$C^{1+\theta,2+\theta}$ regularity for the hydrodynamic equation. This allows us to obtain that $C^{1+\theta,2+\theta}$ solutions
exist and remain in the sub-critical region for all times. So in this particular case, the result on the hydrodynamic limit is
global in time.

Maximum principles are more complicated for non-linear parabolic systems, since one has to determine the shape of the
\emph{invariant region} in which the solution will have to remain~\cite{EvansLC2010a,Wang1990a}, while in the scalar case the
invariant region is just an interval. For the species-blind process we find that the invariant region of the hydrodynamic equation
coincides with the sub-critical region of the ZRP. This is not surprising since the species-blind process is obtained from a
one-species ZRP by colouring particles in two colours, say black and white. The dynamics is the usual ZRP dynamics but at each
time of a jump from a site $x$, we choose the colour of particle to move with the probabilities given by the ratios of the number
of black particles and white particles at $x$ to the total number of particles at $x$, ignoring the colour. It would still be
interesting to study the class of parabolic systems arising from $2$-species ZRPs in order to determine the largest class of ZRPs
that their sub-critical region is an invariant region of the hydrodynamic limit. This would then provide a way to find the
invariant region of the associated parabolic systems by calculating the phase diagram of the underlying ZRP. The study of the
system of PDEs arising from the ZRP is a different topic and outside of the scope of this article, which mainly focuses on the
passage from the microscopic to the macroscopic level by applying the relative entropy method.

\paragraph{Plan of the paper} The paper is organised as follows. In Section~\ref{Model} we collect some preliminary material on
two-species ZRPs and describe the particular case of the species-blind ZRP. Section~\ref{MainResults} contains the statements of
the main results, and in Section~\ref{Proofs} we give the proofs.

\section{The particle model}
\label{Model}

We briefly give the definition of two-species ZRPs as Markov jump processes (Section A.1.2 in~\cite{Kipnis1999a}) and their
equilibrium states. Main references on this preliminary material are~\cite{Grosskinsky2004a,Grosskinsky2008a}. We take the
discrete $d$-dimensional $N$-torus $\T_N^d$ as underlying lattice. Each particle interacts only with particles in the same lattice
site through a function $\bs{g}=(g_1,g_2) \colon \NNZ^2\to\RR_+^2$. Here $g_i(\bs{k})$ is the jump rate of species of type $i$
from any site that contains $\bs{k}\in\NNZ^2$ particles, i.e., $k_i$ particles of type $i$, for $i=1,2$. 
We impose the natural condition
\begin{align}
  \label{NonDegen} 
  g_i(k)=0\quad\text{iff}\quad k_i=0,\qquad\bs{k}=(k_1,k_2)\in\NNZ^2
\end{align}
and require
\begin{align}
  \label{PointWLip}
  \|\pd_ig_i\|_\infty:=\sup_{\bs{k}\in\NNZ^2}|g_i(\bs{k}+\bs{e}_i)-g_i(\bs{k})|<+\infty,
\end{align} 
where $\bs{e}_i=(\delta_{ij})_{j=1,2}$, $i=1,2$, are the unit vectors in $\RR^2$. Note that setting
$\bs{g}^*:=\|\pd_1g_1\|_\infty\mx\|\pd_2g_2\|_\infty<+\infty$, we have by~\eqref{NonDegen} and~\eqref{PointWLip} that
\begin{align}
  \label{quasiLipschitz1}
  |\bs{g}(\bs{k})|_p\leq\bs{g}^*|\bs{k}|_p,\text{ for every }\bs{k}\in\NNZ^2,
\end{align}
where $|\cdot|_p$ denotes the $\ell_p$-norm in $\RR^2$, $p\in[1,+\infty]$.

The state space of a two-species ZRP consists of all configurations $\bs{\eta}=(\eta_1,\eta_2) \colon \T_N^d\to\NNZ^2$, so that
$\eta_i(x)$ is the number of $i$-type particles at site $x$, for $i=1,2$. For any measurable space $M$ we denote by $\PP(M)$ the
set of all probability measures on $M$. We write $p\in\PP(\ZZ^d)$ for the nearest neighbour (n.n.) elementary step distribution
given by
\begin{equation*}
  p(x):=\fr{2d}\sum_{j=1}^d\1_{\{-e_j,e_j\}}(x),\quad x\in\ZZ^d,
\end{equation*} and by
$p_N\in\PP(\T_N^d)$ its projection on $\T_N^d$ given by $p_N(x) := p(x + N\ZZ^d)$. Also, given a configuration
$\bs{\eta}\in\MM_N^{d;2}:=(\NN_0^2)^{\T_N^d}$, we will denote by $\bs{\eta}^{i;x;y}$, $i = 1,2$, the configuration resulting from
$\bs{\eta}$ by moving a type-$i$ particle from $x$ to $y$. (If $\eta_i(x) = 0$, then we set
$\bs{\eta}^{i;x;y} = \bs{\eta}$.)  The \emph{two-species n.n. symmetric ZRP with jump rate $\bs{g}$ on the discrete
  torus $\T_N^d:= \{0,1,\ldots,N-1\}^d$} is the unique Markov jump process on the Skorohod space
$D(\RR_+;\MM_N^{d;2})$ of c\`{a}dl\`{a}g paths characterised by the formal generator
\begin{align}
  \label{2speciesGen} 
  L^Nf(\bs{\eta})=\sum_{i=1,2}\sum_{x,y\in\T_N^d}\{f(\bs{\eta}^{i:x,y})-f(\bs{\eta})\}g_i(\bs{\eta}(x))p_N(y-x).
\end{align} 
We will denote by $(P_t^N)_{t\geq 0}$ the transition semigroup of the n.n. symmetric ZRP.  The communication classes of the
stochastic dynamics defined by the generator above are the hyperplanes
\begin{equation*}
  \MM_{N,\bs{K}}^{d;2}:=\Big\{\bs{\eta}\in\MM_N^{d;2} \bigm|\sum_{x\in\T_N^d}\bs{\eta}(x)=\bs{K}\Big\}
\end{equation*} 
consisting of a fixed number of particles of each species. Since each set $\MM_{N,K}^{d;2}$ is finite, for each
$(N,\bs{K})\in\NN\x\NNZ^2$ there exists a unique equilibrium distribution $\nu_{N,\bs{K}}$ supported on
$\MM_{N,\bs{K}}^{d;2}$. The family $\{\nu_{N,\bs{K}}\}_{(N,\bs{K})\in\NN\x\NNZ^2}$ is the so-called {\emph{canonical
    ensemble}}. However, as proved in~\cite[Theorem 4.1]{Grosskinsky2004a}, in order to have product and translation invariant
equilibrium distributions, it is necessary and sufficient that the following compatibility relations for the component functions
of two-species jump rates hold,
\begin{align}
  \label{CompatibilityCondition}
  g_1(\bs{k})g_2(\bs{k}-\bs{e}_1) = g_1(\bs{k}-\bs{e}_2)g_2(\bs{k}),\quad\text{ for all }\bs{k}\in\NNZ^2\text{ with }k_1, k_2\geq 1.
\end{align}	
Note that due to the compatibility relations~\eqref{CompatibilityCondition} any two-species local jump rate $\bs{g}$ is uniquely
determined by $g_1$ and the restriction of $g_2$ to the set $\{0\}\x\NNZ$, since by induction for any $\bs{k}\in\NNZ^2$
\begin{equation*}
  g_2(\bs{k})=g_2(0,k_2)\prod_{i=0}^{k_1-1}\frac{g_1(\bs{k}-i\bs{e}_1)}{g_1(\bs{k}-i\bs{e}_1-\bs{e}_2)}.
\end{equation*}
An \emph{increasing path $\bs{\g}$ (from $0$) to $\bs{k}\in\NNZ^2$} is any path $\bs{\g}\colon\{0,\ldots,k_1+k_2\}\to\NNZ^2$ such
that $\bs{\g}(0) = 0$, $\bs{\g}(k_1 + k_2)=\bs{k}$ and $\bs{\g}(\ell) = \bs{\g}(\ell-1)+\bs{e}_{i_\ell}$ for some
$i_\ell\in\{1,2\}$ for all $\ell= 1,\ldots,k_1+k_2$. For any increasing path $\g$ to $\bs{k}\in\NNZ^2$, the factorial of $\bs{g}$
along $\bs{\g}$ is defined as
\begin{equation*}
  \bs{g}!(\bs{k};\bs{\g}) =\prod_{\ell=1}^{k_1+k_2}g_{i_\ell}(\bs{\g}(\ell))
\end{equation*}
for $\bs{k}\neq 0$; we set $\bs{g}!(\cdot) := 1$ if $\bs{k}=0$. A two-species local jump rate function $\bs{g}$ that
satisfies~\eqref{CompatibilityCondition} yields a well-defined function $\bs{g}!\colon\NNZ^2\to(0,\infty)$ by the formula
\begin{equation*}
  \bs{g}!(\bs{k}) = \bs{g}!(\bs{k};\bs{\g})\quad\text{ for some increasing path }\bs{\g}\text{ to }\bs{k}.
\end{equation*}
For instance
\begin{align*}\bs{g}!(\bs{k})&= g_1(1,0)\cdot\ldots\cdot g_1(k_1,0)\cdot g_2(k_1, 1)\cdot\ldots\cdot g_2(k_1,k_2)\\
                             &= g_2(0,1)\cdot\ldots\cdot g_2(0,k_2)\cdot g_1(1,k_2)\cdot\ldots\cdot
                               g_1(k_1,k_2).
\end{align*}

According to~\cite[Theorem 4.1]{Grosskinsky2004a}, using the multi-index notation
$\bs{\phih}^{\bs{k}}:=\phih_1^{k_1}\phih_2^{k_2}$, with $\bs{\phih},\bs{k}\in\RR_+^2$, for two-species symmetric n.n.~ZRP
satisfying~\eqref{CompatibilityCondition}, the common one-site marginal $\bar{\nu}_{\bs{\phih}}^1$ of the product and translation
invariant equilibrium states $\bar{\nu}_{\bs{\phih}}^N$ is given by the formula
\begin{equation*}
  \bar{\nu}_{\bs{\phih}}^1(\bs{k})=\fr{Z(\bs{\phih})}\frac{\bs{\phih}^{\bs{k}}}{\bs{g}!(\bs{k})},\quad\bs{k}\in\NNZ^2,
\end{equation*}
for all $\bs{\phih}\in\RR_+^2$ such that the series
\begin{align}
  \label{PartitionFunction} 
  Z(\bs{\phih}):=\sum_{\bs{k}\in\NNZ^2}\frac{\bs{\phih}^{\bs{k}}}{\bs{g}!(\bs{k})}
\end{align}
converges.  The function $Z\colon\RR_+^2\to[0,+\infty]$ defined in~\eqref{PartitionFunction} is called the \emph{partition
  function}. The main convexity property of $Z$ is that the function $\Z:=Z\circ\exp\colon\RR^2\to(-\infty,+\infty]^2$ is strictly
logarithmically convex where $\exp(\bs{\mu}) := e^{\bs{\mu}}:=(e^{\mu_1},e^{\mu_2})$. This can be seen by applying H\"{o}lder's
inequality to the functions $\bs{k}\mapsto e^{\ls\bs{\mu},\bs{k}\rs}$, $\bs{k}\mapsto e^{\ls\bs{\nu},\bs{k}\rs}$ with respect to
the $\s$-finite measure $\lambda$ on $\NNZ^2$ given by $\lambda({\bs{k}}):=\fr{\bs{g}!(\bs{k})}$ and with the pair of conjugate
exponents $p=\fr{1-t}$, $q=\fr{t}$ for $t\in(0,1)$, $\bs{\mu},\bs{\nu}\in\RR^2$, which yields
\begin{equation*}
  \Z\big((1-t)\bs{\mu}+t\bs{\nu}\big)=\int
  e^{(1-t)\ls\bs{\mu},\bs{k}\rs}e^{t\ls\bs{\nu},\bs{k}\rs}d\lambda(\bs{k})\leq\Z(\bs{\mu})^{(1-t)}\Z(\bs{\nu})^t.
\end{equation*}
Here and in what follows $\ls\bs{\mu},\bs{k}\rs=\mu_1k_1+\mu_2k_2$ denotes the Euclidean inner product of two vectors
$\bs{k},\bs{\mu}\in\RR_+^2$. We denote by $\DD_Z:=\{\bs{\phih}\in\RR_+^2\bigm|Z(\bs{\phih})<+\infty\}$ the \emph{proper domain} of
Z, which is a complete, i.e., $[0,\bs{\phih}]:=[0,\phih_1]\x[0,\phih_2]\subs\DD_Z$ for all $\bs{\phih}\in\DD_Z$, and
logarithmically convex set, that, is the set
$\DD_\Z=\log(\DD_Z\cap(0,\infty)^2):=\{\log\bs{\phih}:=(\log\phih_1,\log\phih_2)\,\big|\,\bs{\phih}\in\DD_Z\cap(0,\infty)^2\}$ is
convex. The partition function is $C^\infty$ in $\DD_Z^o$ and continuous from below on $\DD_Z$, i.e., for all
$\bs{\phih}\in\DD_Z$, $\ee>0$ there exists $\delta>0$ such that $|Z(\bs{\phih})-Z(\bs{\psi})|<\ee$ for all $\bs{\psi}\in
D(0,\delta)\cap[0,\bs{\phih}]$. Here $D(0,\delta)$ denotes the Euclidean open ball of radius $\delta$ with centre $0$ in
$\RR_+^2$, i.e., $D(0,\delta)=\big\{\bs{\phih}\in\RR_+^2\bigm||\bs{\phih}|_2<\delta\big\}$.

The family of the product and translation invariant equilibrium states is the family
$\{\bar{\nu}^N_{\bs{\phih}}\}_{\bs{\phih}\in\DD_Z}$. This family is usually referred to as the \emph{grand canonical ensemble
  (GCE)}. In order to ensure that $\DD_Z$ is not trivial, i.e., that $\DD_Z$ contains a neighbourhood of zero in $\RR_+^2$, we
must impose the following condition in the definition of two-species local jump rate functions:
\begin{align}
  \label{NonDegenDom}\qquad
  \phih_{*;1}:=\liminf_{|\bs{k}|_1\ra+\infty}{\bs{g}!(\bs{k})}^{\fr{|\bs{k}|_1}} > 0.
\end{align}
A two-species local jump rate $\bs{g}$ satisfies~\eqref{NonDegenDom} iff $\DD_Z$ contains a neighbourhood of $0$ in $\RR_+^2$. In
what follows, we consider only two-species local jump rates that satisfy~\eqref{CompatibilityCondition} and~\eqref{NonDegenDom}.

It is convenient to have a parametrisation of the GCE by the density. This is done via the density function
$\bs{R} = (R_1,R_2)\colon\DD_Z\to[0,+\infty]^2$ defined by
\begin{equation*}
  \bs{R}(\bs{\phih})=\int_{\MM_N^{d;2}}\bs{\eta}(0)d\bar{\nu}_{\bs{\phih}}^N=\Big(\int k_1d\bar{\nu}_{\bs{\phih}}^1,\int 
  k_2d\bar{\nu}_{\bs{\phih}}^1\Big).
\end{equation*} 
The \emph{proper domain} of $\bs{R}$ is the set $\DD_{\bs{R}}:=\left\{\bs{\phih}\in D_Z\bigm|\bs{R}(\bs{\phih})\in\RR_+^2\right\}$
and by differentiation of bivariate power-series, we have that
\begin{equation}
  \label{DensityFormula}
  \bs{R}(\bs{\phih})=\bs{\phih}\cdot\nabla(\log Z)(\bs{\phih})\quad\text{ on the set }\DD_{\bs{R}}^o=\DD_Z^o,
\end{equation} 
where $\bs{\phih}\cdot\bs{\psi}:=(\phih_1\psi_1,\phih_2\psi_2)$ denotes the pointwise product of two vectors
$\bs{\phih},\bs{\psi}\in\RR_+^2$. Furthermore, this formula extends to the set $\DD_Z\cap\pd\DD_Z$ if we interpret the directional
derivatives $\pd_i(\log Z)\in[0,+\infty]$ as derivatives from the left. With the conventions $\log 0=-\infty$ and $e^{-\infty}=0$
the densities $\bs{\rho}\in\bs{R}(\DD_{\bs{R}})$ can also be parametrised via the chemical potential by the function
$\bs{\R}:=\bs{R}\circ\exp\colon\DD_{\bs{\R}}\to\bs{R}(\DD_{\bs{R}})$, where
$\DD_{\bs{\R}}=\log(\DD_{\bs{R}}):=\{\log\bs{\phih}\in[-\infty,+\infty)^2|\bs{\phih}\in\DD_{\bs{R}}\}$. For the parametrisation
  via the chemical potentials $\bs{\R}(\bs{\mu})=\nabla(\log\Z)(\bs{\mu})$ for all
  $\bs{\mu}\in\DD_{\bs{\R}}^o\cap(-\infty,+\infty)^2$, where $\Z=Z\circ\exp$.

The density function $\bs{R}\colon\bs{R}(\DD_{\bs{R}})\to\DD_{\bs{R}}$ is invertible. Indeed, it is straightforward to check
(e.g., see~\cite[(4.10)]{Grosskinsky2008a}) that for all $\bs{\phih}\in\DD_{\bs{R}}^o\cap(0,+\infty)^2$,
\begin{equation*}
  D\bs{R}(\bs{\phih})=D\bs{\R}(\log\bs{\phih})\begin{pmatrix}\fr{\phih_1}&0\\
0&\fr{\phih_2}\end{pmatrix}={\rm{Cov}}(\bar{\nu}_{\bs{\phih}}^1)\begin{pmatrix}\fr{\phih_1}&0\\
    0&\fr{\phih_2}\end{pmatrix},
\end{equation*}
where ${\rm{Cov}}(\bar{\nu}_{\bs{\phih}}^1)$ denotes the covariance matrix
\begin{equation*}
{\rm{Cov}}(\bar{\nu}_{\bs{\phih}}^1)_{ij}
=\int k_ik_jd\bar{\nu}_{\bs{\phih}}^1-\int k_i\bar{\nu}_{\bs{\phih}}^1\int k_j\bar{\nu}_{\bs{\phih}}^1,\quad i,j=1,2.
\end{equation*} 
This implies that $D\bs{R}(\bs{\phih})$ is diagonisable with strictly positive eigenvalues for all $\bs{\phih}\in
\DD_{\bs{R}}^o\cap(0,+\infty)^2$. Furthermore,
\begin{equation}
  \label{PositDiagElements}
  \pd_1R_1(\bs{\phih})\mn\pd_2R_2(\bs{\phih})>0\mbox{ for all }\bs{\phih}\in\DD_{\bs{R}}^o
\end{equation}
and for $\bs{\phih}\in\DD_{\bs{R}}^o$ such that $\phih_1\phih_2=0$, the matrix $D\bs{R}(\bs{\phih})$ is triangular, and thus
invertible. Therefore the density function $\bs{R}\colon\DD_{\bs{R}}^o\to\bs{R}(\DD_{\bs{R}}^o)$ is invertible. The fact that
$\bs{R}$ is invertible on all of its domain follows by~\cite[Proposition 2.3]{Grosskinsky2008a}, according to which for every
$\bs{\rho}\in(0,\infty)^2$ there exists a unique maximiser $\bar{\bs{\Phi}}(\bs{\rho})\in\DD_{\bs{R}}\cap(0,\infty)^2$ for the
thermodynamic entropy
\begin{align}
  \label{PhiExtension}
  S(\bs{\rho}):= \sup_{\bs{\phih}\in\DD_Z\cap(0,\infty)^2}\{\ls\bs{\rho},\log\bs{\phih}\rs-\log 
  Z(\bs{\phih})\}=\ls\bs{\rho},\log\bar{\bs{\Phi}}(\bs{\rho})\rs-\log Z\big(\bar{\bs{\Phi}}(\bs{\rho})\big).
\end{align} 
Obviously, for $\bs{\rho}=0$ the supremum is attained at $\bs{\phih}=0$ (with the convention $0\cdot(-\infty)=0$). Furthermore,
since $Z$ is non-decreasing with respect to each variable separately, for any $\bs{\rho}\in\RR_+^2\sm\{0\}$ with $\rho_1\rho_2=0$,
say $\rho_2=0$, the maximisation problem~\eqref{PhiExtension} is reduced to the corresponding maximisation problem for one of the
$1$-species jump rate $\hat{g}_1(k):=g_1(k,0)$, $k\in\NNZ$ and the supremum is attained at
$\bar{\bs{\Phi}}(\rho_1,0)=(\hat{\Phi}_1(\rho_1\mn\hat{\rho}_{c,1}),0)$, where $\hat{\Phi}_1$, $\hat{\rho}_{c,1}$ are the mean
jump rate and critical density of the $1$-species jump rate $\hat{g}_1$ (see~\cite[Section 5.2.1]{Grosskinsky2004a} for the
$1$-species case). Thus for any $\bs{\rho}\in\RR_+^2$ there exists a unique maximiser $\bar{\bs{\Phi}}(\bs{\rho})\in \DD_{\bs{R}}$
for the thermodynamic entropy $S(\bs{\rho})$. As in~\cite[Proposition 2.3]{Grosskinsky2008a} the function
$\bs{\Phi}\colon\RR_+^2\to\DD_{\bs{R}}$ is continuous, $\bs{\Phi}:=\bar{\bs{\Phi}}|_{\bs{R}(\DD_{\bs{R}})}=\bs{R}^{-1}$ is the
inverse of $\bs{R}\colon\DD_{\bs{R}}\to\bs{R}(\DD_{\bs{R}})$ and
\begin{equation*}
  \bar{\bs{\Phi}}\big(\RR_+^2\sm\bs{R}(\DD_{\bs{R}})\big)=\DD_{\bs{R}}\cap\pd\DD_{\bs{R}}.
\end{equation*}
Furthermore $\bs{R}(\DD_{\bs{R}})$ is closed in $\RR_+^2$ and
$\pd\bs{R}(\DD_{\bs{R}})=\bs{R}(\DD_{\bs{R}}\cap\pd\DD_{\bs{R}})$. According to this result
$\bar{\bs{\Phi}}\colon\RR_+^2\to\DD_{\bs{R}}$ is a left inverse for $\bs{R}$, i.e., $\bar{\bs{\Phi}}\circ\bs{R} =
\bs{\Phi}\circ\bs{R} = \mathbbm{id}_{\DD_{\bs{R}}}$ and the function
\begin{equation*}
{\bs{R}}_c := \bs{R}\circ\bar{\bs{\Phi}}\colon\RR_+^2\to\bs{R}(\DD_{\bs{R}})
\end{equation*}
is a continuous projection on $\bs{R}(\DD_{\bs{R}})$ with
$\bs{R}_c|_{\bs{R}(\DD_{\bs{R}})} = \mathbbm{id}_{\bs{R}(\DD_{\bs{R}})}$, satisfying 
\begin{equation*}
\bs{R}_c\big(\RR_+^2\sm\bs{R}(\DD_{\bs{R}})\big) = \bs{R}\big(\DD_{\bs{R}}\cap\pd\DD_{\bs{R}}\big) = \pd\bs{R}(\DD_{\bs{R}}).
\end{equation*}
In particular, $\bs{R}\colon\DD_{\bs{R}}\to\bs{R}(\DD_{\bs{R}})$ is a homeomorphism and
$\bs{R}(\DD_{\bs{R}})^o=\bs{R}(\DD_{\bs{R}}^o)$.

Note that the thermodynamic entropy coincides with the Legendre transform of the convex \emph{thermodynamic pressure}
$\log\Z\colon \RR^2\to[0,+\infty]$, that is, 
\begin{equation}\label{ThEnt}
S(\bs{\rho})=(\log\Z)^*(\bs{\rho})=\sup_{\bs{\mu}\in\RR^2}\big\{\ls\bs{\rho},\bs{\mu}\rs-\log\Z(\bs{\mu})\big\}.
\end{equation} Since $\nabla(\log\Z)=\bs{\R}=\bs{R}\circ\exp$, it follows by the formula for the derivative of the Legendre transforms that for all
$\bs{\rho}\in(0,\infty)^2\cap\bs{R}(\DD_{\bs{R}}^o)$ the supremum in~\eqref{ThEnt} is attained at
\begin{equation*} 
\nabla S(\bs{\rho})=(\nabla\log\Z)^{-1}(\bs{\rho})=\bs{\R}^{-1}(\bs{\rho})=\log\bs{\Phi}(\bs{\rho}).
\end{equation*}
Since $S$ is convex the matrix $D^2S(\bs{\rho})=D(\log\bs{\Phi})(\bs{\rho})$ is symmetric and strictly positive definite for all
$\bs{\rho}\in\bs{R}(\DD_{\bs{R}}^o)\cap(0,+\infty)^2$. The symmetry of $D^2S(\bs{\rho})$,
$\bs{\rho}\in\bs{R}(\DD_{\bs{R}}^o)\cap(0,+\infty)^2$, implies the relations
\begin{equation}
  \label{MacrCompCond}
  \Phi_2(\bs{\rho})\pd_2\Phi_1(\bs{\rho})=\Phi_1(\bs{\rho})\pd_1\Phi_2(\bs{\rho}),
\end{equation} 
which extend to $\bs{\rho}\in\bs{R}(\DD_{\bs{R}}^o)$ because $R_i(\bs{\phih})=0$ if and only if $\phih_i=0$, $i=1,2$ and
$D\bs{R}(\bs{\phih})$ is triangular for $\bs{\phih}\in\DD_{\bs{R}}^o$ with $\phih_1\phih_2=0$. Equation~\eqref{MacrCompCond} can
be seen as the macroscopic analogue of the compatibility relations~\eqref{CompatibilityCondition}.

Using the inverse $\bs{\Phi}$ of $\bs{R}$ on $\bs{R}(\DD_{\bs{R}})$, we can parametrise the grand canonical measures
$\bar{\nu}^N_{\bs{\phih}}$, $\bs{\phih}\in\DD_{\bs{R}}$, that have finite density via
\begin{align}
\label{GCE}
\nu_{\bs{\rho}}^N:=\bar{\nu}_{{\bs{\Phi}}(\bs{\rho})}^N,\quad\bs{\rho}\in\bs{R}(\DD_{\bs{R}}),
\end{align}
so that they are parametrised by their density. We will denote by
$\nu^\infty_{\bs{\rho}}:=\bigotimes_{x\in\ZZ^d}\nu^1_{\bs{\rho}}$, $\bs{\rho}\in\bs{R}(\DD_{\bs{R}})$, the product measures on the
configuration space $\MM_\infty^{d;2}:=(\NNZ^2)^{\ZZ^d}$ over the infinite lattice $\ZZ^d$. The \emph{logarithmic
  moment-generating function} $\Lambda_{\bs{\rho}}:=\Lambda_{\nu_{\bs{\rho}}^1}\colon\RR^2\to(-\infty,+\infty]$ of the one-site
marginal $\nu_{\bs{\rho}}^1$, $\bs{\rho}\in\DD_{\bs{R}}$, is defined by
\begin{align}
  \label{LMGF}
  \Lambda_{\bs{\rho}}^1(\bs{\lambda}):=\log\int e^{\ls\bs{\lambda},\bs{k}\rs}d\nu_{\bs{\rho}}^1(\bs{k})
  =\log\frac{Z(e^{\bs{\lambda}}\cdot\bs{\Phi}(\bs{\rho}))}{Z(\bs{\Phi}(\bs{\rho}))}.
\end{align}
Consequently, the product and translation invariant equilibrium states have some exponential moments for all
$\bs{\rho}\in\bs{R}(\DD_{\bs{R}}^o)$. They have full exponential moments iff $\DD_Z=\RR_+^2$.

It is easy to verify that $\bs{\Phi}(\bs{\rho})$ has a probabilistic interpretation as the one-site mean jump rate with respect to
the product and translation invariant equilibrium state of density $\bs{\rho}\in\bs{R}(\DD_{\bs{R}})$, that is
\begin{equation*}
  \bs{\Phi}(\bs{\rho})=\int\bs{g}(\bs{\eta}(0))d\nu_{\bs{\rho}}^N,\quad\bs{\rho}\in\bs{R}(\DD_{\bs{R}}).
\end{equation*}
Since $\bar{\bs{\Phi}}=\bs{\Phi}\circ\bs{R}_c$, it follows by~\eqref{quasiLipschitz1} that for all $\bs{\rho}\in\RR_+^2$
\begin{align}
  \label{PhiQuasiLip}
  \big|\bar{\bs{\Phi}}(\bs{\rho})\big|_1\leq\int\big|\bs{g}(\bs{\eta}(0))
  \big|_1d\nu_{\bs{R}_c(\bs{\rho})}^N\leq\bs{g}^*\int|\bs{\eta}(0)|_1d\nu_{\bs{R}_c(\bs{\rho})}^N
  =\bs{g}^*|\bs{R}_c(\bs{\rho})|_1\leq\bs{g}^*|\bs{\rho}|_1.
\end{align} 

One says that \emph{the $2$-species ZRP is condensing} when $\bs{R}(\DD_{\bs{R}})\neq\RR_+^2$, in which case there exist densities
$\bs{\rho}\in\RR_+^2$ for which there is no grand canonical equilibrium state of density $\bs{\rho}$. Since $\bs{R}(\DD_{\bs{R}})$
is non-empty and closed in $\RR_+^2$ it follows that $\bs{R}(\DD_{\bs{R}})\neq\RR_+^2$ if and only if
$\pd\bs{R}(\DD_{\bs{R}})\neq\emptyset$, and thus condensation occurs precisely when
$\DD_{\bs{R}}\cap\pd\DD_{\bs{R}}\neq\emptyset$. By~\cite[Theorem 3.3]{Grosskinsky2008a} it follows that
$\bs{R}_c(\bs{\rho})\leq\bs{\rho}$, that is, $R_{c,i}(\bs{\rho}):=R_i(\bar{\bs{\Phi}}(\bs{\rho}))\leq\rho_i$, $i=1,2$, for all
$\bs{\rho}\in\RR^2$. One says that \emph{condensation of the $i$-th species, $i=1,2$, occurs at the density $\bs{\rho}\in\RR_+^2$}
if $R_{c,i}(\bs{\rho}) < \rho_i$. All cases are possible, that is, at a given density $\bs{\rho}\in\RR_+^2$ no condensation,
condensation of exactly one species and condensation of both species simultaneously can occur. These cases induce an obvious
partition of the phase space $\RR_+^2$.

As proved in~\cite{Grosskinsky2008a}, the extension $\bar{\bs{\Phi}}$ is the correct one for the equivalence of ensembles in the
sense that $\bs{R}_c$ gives the correct limiting background density in the thermodynamic limit. In the case of condensation, i.e.,
when $\bs{R}(\DD_{\bs{R}})\neq\RR_+^2$, some additional assumption must be imposed on the jump rate $\bs{g}$ to ensure that for
each $\bs{\phih}\in\DD_{\bs{R}}\cap\pd\DD_{\bs{R}}$, the one-site marginal $\bar{\nu}_{\bs{\phih}}^1$ has heavy tails in the
direction normal to the set
$\log(\DD_{\bs{R}}\cap\pd\DD_{\bs{R}}):=\{\log\bs{\phih}|\bs{\phih}\in\DD_{\bs{R}}\cap\pd\DD_{\bs{R}}\}$ at
$\bs{\mu}:=\log\bs{\phih}$. Denoting by $n_{\bs{\phih}}$ the normal to $\log(\DD_{\bs{R}}\cap\pd\DD_{\bs{R}})$ at $\log\bs{\phih}$
(where $n_{(\phih_1,0)}=\bs{e}_1$, $n_{(0,\phih_2)}=\bs{e}_2$), this means that
\begin{align}
  \label{SubExpTails}
  \lim_{\substack{|\bs{k}_n|_2\ra+\infty\\{\bs{k}_n}/{|\bs{k}_n|_2}\ra n_{\bs{\phih}}}}\fr{|\bs{k}_n|_2}\log\bar{\nu}_{\bs{\phih}}^1(\bs{k}_n)=0.
\end{align} 
In case $\pd\DD_{\bs{R}}$ is not differentiable at $\bs{\phih}$, \eqref{SubExpTails} is required to hold for the two limiting
normal vectors $n_{\bs{\phih}}^+$, $n_{\bs{\phih}}^-$ at $\log\bs{\phih}$. As has been proven in~\cite[Lemma
3.5]{Grosskinsky2008a}, a condition on the jump rate $\bs{g}$ that guarantees the critical equilibrium states have heavy tails in
the direction normal to the logarithm of the boundary is the \emph{regularity of its tails}, in the sense that for any direction
$\bs{\y}\in S^1_+:=S^1\cap\RR_+^2$,
\begin{align}
  \label{RegTails}
  \phih_{c;2}(\bs{\y}):=\liminf_{\substack{|\bs{k}|_2\ra+\infty\\ {\bs{k}}/{|\bs{k}|_2}\ra \bs{\y}}}\bs{g}!(\bs{k})^\fr{|\bs{k}|_2}\in(0,\infty)
\end{align}
exists as limit and $\phih_{c;2} \colon S^1_+\to(0,\infty)$ is a continuous function of the direction $\bs{\y}\in S^1_+$.  Note
that instead of the exponent $p=2$, we could have used any $p\in[1,+\infty]$, replacing the Euclidean sphere $S^1_+$ with the
sphere $S^1_{p,+}:=\{\bs{x}\in\RR_+^2 \bigm| |\bs{x}|_p=1\}$ with respect to the $\ell_p$-norm on $\RR_+^2$. According to the
equivalence of ensembles~\cite[Theorem 3.1]{Grosskinsky2008a}, if the jump rate has regular tails when
$\bs{R}(\DD_{\bs{R}})\neq\RR_+^2$, then for all $\bs{\rho}\in\RR_+^2$
\begin{align}
  \label{EqOfEns}  
  \lim_{\substack{N,|\bs{K}|\ra+\infty\\\bs{K}/N^d\ra\bs{\rho}}}\fr{N^d}\HHH(\nu_{N,\bs{K}}|\nu_{\bs{R}_c(\bs{\rho})}^N)=0.
\end{align}
Here $\HHH(\mu|\nu)$ denotes the \emph{relative entropy} between two probability measures $\mu,\nu$,
\begin{equation*}
  \HHH(\mu|\nu):=
  \begin{cases}\int\frac{d\mu}{d\nu}\log\frac{d\mu}{d\nu}d\nu\quad&\text{if }\mu\ll\nu\\
    +\infty\quad&\text{otherwise}
  \end{cases}.
\end{equation*}
The translation invariance of canonical and grand canonical ensembles and the super-additivity of the relative entropy imply
convergence for any finite set $F\subs\ZZ^d$, i.e.,
\begin{equation*}
  \lim_{\substack{N,|\bs{K}|\ra+\infty\\\bs{K}/N^d\ra\bs{\rho}}}\HHH(\nu_{N,\bs{K}}^F|\nu_{\bs{R}_c(\bs{\rho})}^{N,F})=0
\end{equation*}
where $\nu_{N,\bs{K}}^F:=p_{F*}\nu_{N,\bs{K}}$, $\nu_{\bs{R}_c(\bs{\rho})}^{N,F}:=p_{F*}\nu_{\bs{R}_c(\bs{\rho})}^N$ are the
push-forwards via the natural projection $p_F\colon\MM_N^{d;2}\to(\NNZ^2)^F$ and $\T_N^d$ is considered embedded in $\ZZ^d$. In
turn this implies that $\nu_{N,\bs{K}}$ (considered embedded in the larger space $\MM_\infty^{d;2}$) converges as
$\bs{K}/N^d\to\bs{\rho}$ to $\nu_{\bs{R}_c(\bs{\rho})}^\infty$ weakly with respect to bounded \emph{cylinder functions}
$f\colon\MM_\infty^d\to\RR$, that is, such that they depend on a finite number of coordinates.

Finally, we briefly recall the notions of local equilibrium and hydrodynamic limits and refer to~\cite{Kipnis1999a} for more
details. We say that a sequence of probability measures $\{\mu^N\in\PP(\MM_N^{d;2})\}$ is an \emph{entropy-local equilibrium of
  profile $\bs{\rho}\in C(\T^d;\bs{R}(\DD_{\bs{R}}))$} if
\begin{align}
  \label{LittleEntrAss}
  \limsup_{N\ra+\infty}\fr{N^d}\HHH(\mu^N|\nu_{\bs{\rho}(\cdot)}^N)=0.
\end{align} 
Here $\nu_{\bs{\rho}(\cdot)}^N:=\bigotimes_{x\in\T_N^d}\nu_{\bs{\rho}(x/N)}^1$ is the \emph{product measure with slowly varying
  parameter associated to the profile} $\bs{\rho}\in C(\T^d;\bs{R}(\DD_{\bs{R}}))$. Given any cylinder function
$f\colon\MM_N^{d;2}\to\RR$, we set $\wt{f}(\bs{\rho}):=\int fd\nu_{\bs{R}_c(\bs{\rho})}^N$, $\bs{\rho}\in\RR_+^2$. By a simple
adaptation of~\cite[Corollary 6.1.3]{Kipnis1999a}, if $\{\mu^N\}$ is an entropy-local equilibrium of profile $\bs{\rho}\in
C(\T^d;\bs{R}(\DD_{\bs{R}}))$, then
\begin{equation}
  \label{LocEq}
  \lim_{N\ra+\infty}\EE_{\mu^N}\Big|\fr{N^d}\sum_{x\in\T_N^d}H\Big(\frac{x}{N}\Big)\tau_xf(\bs{\eta})
  -\int_{\T^d}H(u)\wt{f}\big(\bs{\rho}(u)\big)du\Big|=0
\end{equation}
for all $H\in C(\T^d)$ and all bounded cylinder functions $f\colon\MM_N^{d;2}\to\RR$, that is, $\mu^N$ is a \emph{weak local equilibrium
  of profile $\bs{\rho}\in C(\T^d;\bs{R}(\DD_{\bs{R}}))$.}

The \emph{hydrodynamic limit (in the diffusive timescale $t\mapsto tN^2$)} of the n.n. two-species ZRP is an evolutionary PDE,
such that entropy-local equilibria are \emph{conserved along its solutions (in the diffusive time-scale)} in the following sense:
If we start the process from an entropy local equilibrium $\mu_0^N\in\PP(\MM_N^{d;2})$, $N\in\NN$, of some sufficiently regular
initial profile $\bs{\rho}_0\colon\T^d\to\RR_+^2$ at time $t=0$ and if there exists a sufficiently regular solution $\bs{\rho}$ of
the hydrodynamic equation on $[0,T)\x\T^d$ starting from $\bs{\rho}_0$, then $\mu_t^N:=\mu_0^NP_{tN^2}^N$ is an entropy-local
equilibrium of profile $\bs{\rho}(t,\cdot)$ for each $t\in[0,T)$.

The main goal of this article is to apply the relative entropy method of H.T.~Yau in order to prove the hydrodynamic limit of
condensing $2$-species ZRPs that start from an initial entropy-local equilibrium $\{\mu_0^N\}$ of sub-critical and strictly
positive profile $\bs{\rho}_0\in C(\T^d;\bs{R}(\DD_{\bs{R}}^o)\cap(0,\infty)^2)$, which is stated as Theorem~\ref{HL} below. A
main ingredient in the proof of the hydrodynamic limit is the one-block estimate which is stated as Theorem~\ref{OBE}. The
relative entropy method also requires the existence of a $C_{\rm{loc}}^{1,2+\theta}$ classical solution
$\bs{\rho}\colon[0,T)\x\T^d\to\bs{R}(\DD_{\bs{R}}^o)\cap(0,+\infty)^2$ for the hydrodynamic limit and applies the Taylor expansion
for $C^{2+\theta}$ functions to the function $\bs{\Phi}(\bs{\rho}_t)$ of the solution $\bs{\rho}_t$ at each time $t>0$
(see~\eqref{Expl2}) in order to estimate the entropy production $\pd_t\HHH(\mu_t^N|\nu_{\bs{\rho}_t(\cdot)}^N)$. The
sub-criticality of the solution $\bs{\rho}$, i.e., that $\bs{\rho}([0,T)\x\T^d)\subs\bs{R}(\DD_{\bs{R}}^o)$, is used in
Lemma~\ref{lemmaLitlleoToBigOEntrAssumpt} and to obtain the bound~\eqref{UnifBoundOnGOnFinTImeHorizon}, which is essential in the
application of Lemma~\ref{LastBound}. The sub-criticality of the solution $\bs{\rho}$ is also required for the application of the
Large Deviations Lemma~\ref{LDPlemma}. Together with the $C^{2+\theta}$ regularity of $\bs{\rho}_t$ for each $t\geq 0$ it is the
main assumption on the solution $\bs{\rho}$. Furthermore, in the Taylor expansion the quantities $\Phi_i(\bs{\rho}_t)$, $i=1,2$,
appear in the denominator, so we have to assume that the solution $\bs{\rho}$ is coordinate-wise strictly positive.

    As already mentioned in the introduction the expected hydrodynamic limit of the $2$-species ZRP with product measures is a
    quasilinear parabolic system of the form~\eqref{NonLinDiffSystem}, which in divergence form is given by
\begin{equation}\label{NonLinDiffSystemDivForm}
\pd_t\bs{\rho}=\dv\bs{\A}_{\bs{\Phi}}(\bs{\rho},\nabla\bs{\rho}).
\end{equation}
Here the divergence with respect to the spatial parameter is applied coordinate-wise, and
$\nabla\bs{\rho}(t,u):=(\nabla\rho_1(t,u),\nabla\rho_2(t,u))\in\RR^{2\x d}$ is the gradient of $\bs{\rho}$ with respect to the
spatial variable $u\in\T^d$. Furthermore,
$\bs{\A}_{\bs{\Phi}}=(A_{\bs{\Phi}}^1,\A_{\bs{\Phi}}^2)\colon\bs{R}(\DD_{\bs{R}}^o)\x\RR^{2\x d}\to\RR^{2\x d}$ is the function given by
\begin{equation*}
  \bs{\A}_{\bs{\Phi}}(\bs{\rho},\bs{V})=D\bs{\Phi}(\bs{\rho})\bs{V},
\end{equation*}
that is,
\begin{equation*}
  \pd_t\rho_i=\dv\A^i_{\bs{\Phi}}(\bs{\rho},\nabla\bs{\rho})=\dv\big(\nabla\Phi_i(\bs{\rho})\nabla\bs{\rho}\big)
  =\Delta\Phi_i(\bs{\rho}),\quad i=1,2.
\end{equation*}
Structural properties of the \emph{mobility matrix} $D\bs{\Phi}\colon\bs{R}(\DD_{\bs{R}}^o)\to\RR^{2\x2}$ can be inferred by the
properties of $D\bs{R}$. For example, for all $\bs{\rho}\in\bs{R}(\DD_{\bs{R}}^o)\cap(0,+\infty)^2$,
\begin{equation*}
  D\bs{\Phi}(\bs{\rho})=\begin{pmatrix}\Phi_1(\bs{\rho})&0\\
  0&\Phi_2(\bs{\rho})\end{pmatrix}D^2S(\bs{\rho}),
\end{equation*} 
where $D^2S(\bs{\rho})=D(\log\bs{\Phi})(\bs{\rho})$ is a strictly positive definite matrix, the second derivative of the
thermodynamic entropy, and for all $\bs{\rho}\in\bs{R}(\DD_{\bs{R}}^o)$, the relations~\eqref{MacrCompCond} hold and
\begin{equation*}
  \pd_1\Phi_1(\bs{\rho})\mn\pd_2\Phi_2(\bs{\rho})>0.
\end{equation*}
In particular, $D\bs{\Phi}(\bs{\rho})$ has positive eigenvalues for all $\bs{\rho}\in\bs{R}(\DD_{\bs{R}}^o)$ and is diagonisable
for all $\bs{\rho}\in\bs{R}(\DD_{\bs{R}}^o)\cap(0,+\infty)^2$. For $\bs{\rho}\in\bs{R}(\DD_{\bs{R}}^o)$ with $\rho_1\rho_2=0$, the
matrix $D\bs{\Phi}(\bs{\rho})$ is triangular. It follows that although $D\bs{\Phi}(\bs{\rho})$ is not necessarily symmetric, it is
uniformly parabolic away from the critical densities, that is, for any compact $K\subs\bs{R}(\DD_{\bs{R}}^o)$ the exists
$\lambda_K>0$ such that
\begin{equation*}
  \ls\bs{\xi},D\bs{\Phi}(\bs{\rho})\bs{\xi}\rs\geq\lambda_K|\bs{\xi}|^2,\quad\bs{\rho}\in K,\;\bs{\xi}\in\RR^2.
\end{equation*}

By the work~\cite{Amann1986a} of Amann, which covers uniformly parabolic systems in general form, it is known that for initial
data $\bs{\rho}_0\in C^{2+\theta}(\T^d;\bs{R}(\DD_{\bs{R}}^o)\cap(0,+\infty)^2)$ there exists a unique maximal classical
$C_{\rm{loc}}^{1,2+\theta}$ solution $\bs{\rho}\colon[0,T_{\rm{max}})\x\T^d\to\RR^2$ of~\eqref{NonLinDiffSystemDivForm}, which by
  continuity will remain in $\DD_{\bs{R}}^o\cap(0,+\infty)^2$ on a possibly even smaller time interval. This establishes the local
  in time existence of $C^{1,2+\theta}$ sub-critical solutions $\bs{\rho}$. On the other hand, by the regularity theory of
  quasilinear uniformly parabolic systems of the form~\eqref{NonLinDiffSystemDivForm}, see \cite[Theorem 1.2]{Duzaar2005a} and the
  references therein, it is known that weak solutions to such systems exhibit singularities on a closed subset $Q\subs[0,T]\x\T^d$
  of zero measure. So we can not simply apply the $C^{2+\theta}$ Taylor expansion on the function $\bs{\Phi}(\bs{\rho}_t)$ for all
  times $t\geq 0$. Furthermore we do not know whether the sub-critical region $\bs{R}(\DD_{\bs{R}})$ is an invariant region for
  the $2$-species ZRP system~\eqref{NonLinDiffSystemDivForm}. These are the two main reasons that force us to rely on Amann's
  local in time existence of regular solutions, and prove a local in time version of the hydrodynamic limit. A further study of
  the PDE system arising as the hydrodynamic limit of a $2$-species ZRP, although interesting, is outside of the scope of this
  article, which is the passage from the microscopic to the macroscopic description.

  However, in the example of the species-blind ZRP one can take into advantage its relation with a particular $1$-species ZRP to
  obtain the global in time existence of $C_{\rm{loc}}^{1+\theta,2+\theta}$ solutions and a type of maximum principle, in which
  the sub-critical region plays the role of the invariant domain. We prove this in Theorem \ref{PDE}.

\subsection{The Species-blind ZRP} We now consider two-species local jump rate functions of the form
\begin{equation}
  \label{SpecBlindJumpRate} 
  g_1(\bs{k}) = k_1h(k_1 + k_2),\quad g_2(\bs{k}) = k_2h(k_1 + k_2)
\end{equation}
for some function $h\colon \NNZ\to\RR_+$ satisfying the non-degeneracy condition $h(k)>0$ for all $k\in\NN$. Any jump rate
$\bs{g}$ of this form satisfies~\eqref{CompatibilityCondition} since
\begin{equation*}
  g_1(\bs{k})g_2(\bs{k}-\bs{e}_1) = k_1h(k_1 + k_2)k_2h(k_1 + k_2 - 1) = g_1(\bs{k} - \bs{e}_2)g_2(\bs{k})
\end{equation*}
for all $\bs{k}\in\NN^2$ and the factorial of such a jump rate is given by
\begin{equation*}
  \bs{g}!(\bs{k})=1\cdot h(1)\cdot\ldots\cdot k_1\cdot h(k_1)\cdot 1\cdot h(k_1 + 1)\cdot\ldots\cdot k_2\cdot h(k_1 + k_2)
  =k_1!k_2!h!(k_1 + k_2).
\end{equation*}
The partition function associated to $\bs{g}$ is given for $\bs{\phih}\in\NNZ^2$ with $\phih_2 > 0$ by
\begin{equation*}
  Z(\bs{\phih})=\sum_{m=0}^\infty\frac{\phih_2^m}{h!(m)}\sum_{k_1=0}^m\frac{(\frac{\phih_1}{\phih_2})^{k_1}}{k_1!(m-k_1)!}
  =\sum_{m=0}^\infty\frac{\phih_2^m}{m!h!(m)}\Big(1+\frac{\phih_1}{\phih_2}\Big)^m=\hat{Z}(\phih_1+\phih_2),
\end{equation*}     
where $\hat{Z}$ is the partition function associated to the $1$-species rate function $\hat{g}(k):=kh(k)$. So, in what follows, we
assume that $h$ is of the form $h(k)=\frac{\hat{g}(k)}{k}$, $k\geq 1$, for some $1$-species local jump rate function $\hat{g}$
with regular tails, i.e., such that the limit inferior $\hat{\phih}_c:=\liminf_{k\ra+\infty}\hat{g}!(k)^{\fr{k}}>0$ exists as a
limit. In this case the function $\bs{g}$ defined in~\eqref{SpecBlindJumpRate} is a two-species local jump rate. Indeed, the
non-degeneracy condition~\eqref{NonDegen} and the Lipschitz condition~\eqref{PointWLip} are easy to verify, as we have seen
$\bs{g}$ satisfies the compatibility condition~\eqref{CompatibilityCondition} and obviously
$\DD_Z=\{\bs{\phih}\in\RR_+^2|\phih_1+\phih_2\in\DD_{\hat{Z}}\}$ and
$\DD_{\bs{R}}=\{\bs{\phih}\in\RR_+^2|\phih_1+\phih_2\in\DD_{\hat{R}}\}$, where $\hat{R}(\phih) =\phih(\log\hat{Z})'(\phih)$ is the
density function associated to the $1$-species jump rate $\hat{g}$. In particular $\DD_Z\neq\emptyset$ and thus
also~\eqref{NonDegenDom} holds. We will refer to this nearest neighbour $2$-species ZRP as the \emph{species-blind ZRP
  corresponding to the $1$-species jump rate $\hat{g}$.} The density function corresponding to $\bs{g}$ is given by the formula
\begin{equation*}
 \bs{R}(\bs{\phih})=\Big(\frac{\phih_1\hat{Z}'(\phih_1+\phih_2)}{\hat{Z}(\phih_1+\phih_2)},
 \frac{\phih_2\hat{Z}'(\phih_1+\phih_2)}{\hat{Z}(\phih_1+\phih_2)}\Big)=\frac{\hat{R}(|\bs{\phih}|_1)}{|\bs{\phih}|_1}\bs{\phih}.
\end{equation*} 
We set $\hat{\Phi}:= \hat{R}^{-1}$ and we will compute the inverse $\bs{\Phi}$ of $\bs{R}\colon\DD_{\bs{R}}\to\RR_+^2$ in its image
$\bs{R}(\DD_{\bs{R}})$. Let $\bs{\rho}=\bs{R}(\bs{\phih})$. We have to solve the system
\begin{equation}
  \label{BlindSystem}   	
  \rho_1=\frac{\phih_1\hat{Z}'(\phih_1+\phih_2)}{\hat{Z}(\phih_1+\phih_2)},\quad\rho_2
  =\frac{\phih_2\hat{Z}'(\phih_1+\phih_2)}{\hat{Z}(\phih_1+\phih_2)}
\end{equation} 
for $(\phih_1,\phih_2)$. By adding the two equations we obtain that $\rho_1+\rho_2=\hat{R}(\phih_1+\phih_2)$. In particular
$\rho_1+\rho_2\in\hat{R}(\DD_{\hat{R}})$ for all $\bs{\rho}\in\bs{R}(\DD_{\bs{R}})$ and
$\phih_1+\phih_2=\hat{\Phi}(\rho_1+\rho_2)$. Substituting $\phih_1 + \phih_2$ with $\hat{\Phi}(\rho_1+\rho_2)$ in both equations
in~\eqref{BlindSystem}, we can solve for $(\phih_1,\phih_2)$ to obtain
\begin{equation*}
  \phih_i=\rho_i\frac{\hat{Z}\big(\hat{\Phi}(\rho_1+\rho_2)\big)}{\hat{Z}'\big(\hat{\Phi}(\rho_1+\rho_2)\big)}=
  \rho_i\fr{(\log\hat{Z}'\big(\hat{\Phi}(\rho_1+\rho_2)\big)}=\rho_i\frac{\hat{\Phi}(\rho_1+\rho_2)}{\rho_1+\rho_2},
\end{equation*}
where the last equality above follows from the identity $\hat{R}(\phih) = \phih(\log Z)'(\phih)$ for the 1-species density and
partition functions, since by this identity we have for all $\rho\in(0,\hat{\rho}_c)$ that \begin{equation*}
  \fr{(\log\hat{Z})'\big(\hat{\Phi}(\rho)\big)}=\frac{\hat{\Phi}(\rho)}{\hat{R}(\hat{\Phi}(\rho))}=\frac{\hat{\Phi}(\rho)}{\rho},
\end{equation*}
where $\hat{\rho}_c$ is the corresponding critical density of the $1$-species jump rate $\hat{g}$. Consequently, the inverse
$\bs{\Phi}:=\bs{R}^{-1}\colon\bs{R}(\DD_{\bs{R}})\to\DD_{\bs{R}}$ is given by the formula
\begin{equation}
  \label{BlProcPhi}   	
  \bs{\Phi}(\bs{\rho})=\Big(\rho_1\frac{\hat{\Phi}(\rho_1+\rho_2)}{\rho_1+\rho_2},
  \rho_2\frac{\hat{\Phi}(\rho_1+\rho_2)}{\rho_1+\rho_2}\Big)=\frac{\hat{\Phi}(|\bs{\rho}|_1)}{|\bs{\rho}|_1}\bs{\rho}.
\end{equation}
Thus the expected hydrodynamic equation of the species-blind ZRP is
\begin{align}
  \label{SumSystemPDE}
  \pd_t\rho_i=\Delta\Big(\rho_i\frac{\hat{\Phi}(\rho_1+\rho_2)}{\rho_1+\rho_2}\Big),\quad i=1,2.
\end{align}
Since~\eqref{SumSystemPDE} is the expected hydrodynamic equation of the species-blind ZRP we will refer to it as the
\emph{species-blind parabolic system}. A \emph{classical solution to the species blind system} is a $C^{1,2}$ function
$\bs{\rho}=(\rho_1,\rho_2)\colon[0,T)\x\T^d\to\RR^2$ satisfying~\eqref{SumSystemPDE} with
$0\leq\rho_1(t,u)+\rho_2(t,u)<\hat{\rho}_c$ for all $(t,u)\in[0,T)\x\T^d$. Note that for any solution $\bs{\rho}=(\rho_1,\rho_2)$
of the system~\eqref{SumSystemPDE}, the sum $\rho_1+\rho_2$ satisfies the parabolic equation $\pd_t\rho=\hat{\Phi}(\rho)$
corresponding to the 1-species ZRP of jump rate $\hat{g}(k)=kh(k)$. This remark will allows us to prove the global in time
existence of solutions to the species-blind system. A similar argument was used for two-species simple exclusion processes
in~\cite{Quastel1992a}.

As an example of the nice properties of the species-blind process, we note that the extended mean jump rate
$\bar{\bs{\Phi}}\colon\RR_+^2\to\DD_{\bs{R}}$ of the species-blind process can be computed explicitly and is given by
\begin{equation*}
  \bar{\bs{\Phi}}(\bs{\rho})=\frac{\bar{\hat{\Phi}}(|\bs{\rho}|_1)}{|\bs{\rho}|_1}\bs{\rho},
\end{equation*}
where $\bar{\hat{\Phi}}(\rho)=\hat{\Phi}(\rho\mn\hat{\rho}_c)$, $\rho\geq 0$, is the extended mean jump rate of the $1$-species
ZRP with jump rate $\hat{g}$.

\section{Main Results}
\label{MainResults} 

A main probabilistic ingredient in the proof of the hydrodynamic limit of ZRPs is the so-called one-block estimate, which is well
known under assumptions that exclude condensing ZRPs (e.g.,~\cite[Section 5.4]{Kipnis1999a}). Our first result is a version of the
one-block estimate for condensing ZRPs, i.e., $\bs{R}(\DD_{\bs{R}})\neq\RR_+^2$, under the additional assumptions that the local
jump rate $\bs{g}$ is bounded, has a continuous partition function $Z$, and has regular tails in the sense of~\eqref{RegTails}. We
note that these extra assumptions in the one-block estimate and the hydrodynamic limit below are not required in the
non-condensing case, i.e., when $\bs{R}(\DD_{\bs{R}})=\RR_+^2$. In the case that $\bs{R}(\DD_{\bs{R}})=\RR_+^2$, but
$\DD_Z\neq\RR_+^2$, Theorems~\ref{OBE} and \ref{HL} still hold under the (weaker than boundedness) assumption that $\bs{g}$ has
sub-linear growth at infinity in the sense that
\begin{align}
  \label{SubLinGrowth} 
  \limsup_{|\bs{k}|_1\ra+\infty}\frac{|\bs{g}(\bs{k})|_1}{|\bs{k}|_1}=0.
\end{align} 
In the case that $\bs{R}(\DD_{\bs{R}})=\DD_Z=\RR_+^2$, no extra assumption is required on $\bs{g}$. Given any (cylinder) function
$\bs{f}\colon\MM_N^{d;2}\to\RR^2$ we set
\begin{equation*}
  \bs{f}^\ell:=\fr{(2\ell+1)^d}\sum_{|x|\leq\ell}\tau_x\bs{f},
\end{equation*} 
where $\tau_x\bs{f}(\bs{\eta}):=\bs{f}(\tau_x\bs{\eta})$ and $\tau_x\bs{\eta}(y):=\bs{\eta}(x+y)$ for $x,y\in\T_N^d$.

\begin{theorem}[One-block estimate]
  \label{OBE} 
  Suppose that the ZRP is condensing and that the local jump rate $\bs{g}$ of the ZRP is bounded, has regular tails in the sense
  of~\eqref{RegTails} and its partition function $Z$ is continuous on $\DD_Z\cap\pd\DD_Z$. Then for any sequence of initial
  distributions $\mu_0^N\in\PP(\MM_N^{d;2})$ satisfying the $O(N^d)$-entropy assumption, i.e.,
  \begin{align}
    \label{EntrAssumpt}
    C(\bs{a}):=\limsup_{N\in\NN}\fr{N^d}\HHH(\mu_0^N|\nu_{\bs{a}}^N)<+\infty,
  \end{align}
  for some (and thus for any) $\bs{a}\in\bs{R}(\DD_{\bs{R}}^o)\cap(0,\infty)^2$, it holds that
  \begin{align}
    \label{TimeDependentOneBlockEstimate}
    \lim_{\ell\ra\infty}\limsup_{N\ra\infty}\EE^N\bigg|\int_0^T\fr{N^d}\sum_{x\in\T_N^d}
    \Big\ls\bs{F}\Big(t,\frac{x}{N}\Big),\bs{g}(\bs{\eta}_t(x))-\bar{\bs{\Phi}}
    \big(\bs{\eta}_t(x)^\ell\big)\Big\rs dt\bigg|=0
  \end{align}
  for all functions $\bs{F}\in C([0,T]\x \T^d;\RR^2)$, $T>0$; $\EE^N$ denotes the expectation with respect to the diffusively
  accelerated law of the ZRP starting from $\mu_0^N\in\PP(\MM_N^{d;2})$ and $\bar{\bs{\Phi}}$ is the extension of $\bs{\Phi}$ given
  by~\eqref{PhiExtension}.
\end{theorem}
     	
We note that the extension $\bar{\bs{\Phi}}$ of the mean jump rate is required in the statement of the one-block estimate, because
$\bs{\eta}_t^\ell$ can be outside the domain of sub-critical densities. This is the correct extension due to the equivalence of
ensembles. The proof of this result is given in Subsection~\ref{sec:Proof-Prop} below.
   	    	
Next is the general result regarding the hydrodynamic limit of two-species ZRPs. As noted in the introduction, in order to take
into account condensing ZRPs, we apply the relative entropy method of H.T.~Yau which requires only the one-block estimate and not
the full replacement lemma. But this method relies on the existence of sufficiently regular classical solutions of parabolic
systems which are known to exist only locally in time, and thus the result is local in time, valid for the time interval that the
unique maximal classical solution of~\eqref{NonLinDiffSystem} established in~\cite{Amann1986a} exists. We denote by
$C^{1+a,2+b}([0,T]\x\T^d)$, $a,b\in[0,1)$, the space of all $C^{1,2}$-functions $f\colon [0,T]\x\T^d\to\RR$ such that
$\pd_tf\in C^a([0,T]\x\T^d)$ is $a$-H\"{o}lder continuous and $\pd^2_{ij}f\in C^b([0,T]\x\T^d)$ is $b$-H\"{o}lder continuous,
where $[0,T]\x\T^d$ is equipped with the parabolic metric $d$ given by
\begin{equation*}
  d\big((t,x),(s,y)\big)=(d_{\T^d}(x,y)^2+|t-s|)^\fr{2}.
\end{equation*}
As usual, if $I\subs\RR$ is an interval, then we write $C^{1+a,2+b}_{\rm{loc}}(I\x\T^d)$ for the space of all functions $f$ such
that $f\in C^{1+a,2+b}(J\x\T^d)$ for any compact sub-interval $J\subs I$. This is extended coordinate-wise to vector-valued
functions; given a subset $A\subs\RR^2$, we denote by $C^{1+a,2+b}_{\rm{loc}}(I\x\T^d;A)$ the subset of
$C^{1+a,2+b}_{\rm{loc}}(I\x\T^d;\RR^2)$ consisting of functions taking values in $A$.

\begin{theorem}[Hydrodynamic limit]
  \label{HL} 
  Let $(S_t^N)_{t\geq 0}$ be the transition semigroup of the two-species symmetric n.n.~ZRP on the torus $\T_N^d$, $N\in\NN$, with
  condensing jump rate $\bs{g}$ satisfying the assumptions of the one-block estimate above, and let $\bs{\Phi}$ be the mean jump
  rate associated to $\bs{g}$. Let
  \begin{equation*}
    \bs{\rho}\in
    C^{1,2+\theta}_{\rm{loc}}\big([0,T_{\rm{max}})\x\T^d;\bs{R}(\DD_{\bs{R}}^o)\cap(0,\infty)^2\big)
  \end{equation*} 
  be the unique maximal solution of the problem~\eqref{NonLinDiffSystem}. Then any initial entropy local equilibrium
  $\mu_0^N\in\PP(\MM_N^{d;2})$ is conserved along the solution $\bs{\rho}$. In other words, if $\{\mu_0^N\}$ is an entropy-local
  equilibrium of profile $\bs{\rho}_0:=\bs{\rho}(0,\cdot)\in C^{2+\theta}(\T^d)$ then $\mu_t^N:=S_{tN^2}^N\mu_0^N$, with
  $N\in\NN$, is an entropy local equilibrium for all $t\in[0,T_{\rm{max}})$. In particular, $\{\mu_t^N\}$ satisfies~\eqref{LocEq}
    for all $t\in[0,T_{\rm{max}})$.
\end{theorem}     	
   
This theorem is proved in Subsection~\ref{sec:Proof-Theor-1}. We should note that, although the proof of the hydrodynamic limit
relies strongly on the assumption that the classical solution $\bs{\rho}$ takes values in the set
$\bs{R}(\DD_{\bs{R}}^o)\cap(0,\infty)^2$ for all times $t\geq 0$, and so in particular requires the sequence of initial
distributions $\{\mu_0^N\}$ to be an entropy local equilibrium of some sub-critical and strictly positive profile
$\bs{\rho}_0\equiv\bs{\rho}(0,\cdot)$, the one-block estimate does not require this assumption. It only requires that
$\{\mu_0^N\}$ satisfies the $O(N^d)$-entropy assumption, which can hold even for super-critical profiles, having a Dirac mass of
order $O(N^d)$ at some site $x\in\T^d$, e.g., $\mu_0^N=\delta_{[\bs{a} N^d]}\otimes\bigotimes_{y\neq[Nx]}\nu_{\bs{\rho}(y/N)}^1$
with $\bs{a}\in(0,\infty)^2$, when $\bs{R}(\DD_{\bs{R}})\neq\RR_+^2$.

We note also that the assumption that $\bs{\rho}([0,T_{\rm{max}})\x\T^d)\subs(0,+\infty)^2$ is a technical one, arising from the
  fact the $\bs{\Phi}(\bs{\rho}_t)$ appears in the denominator. If one knew that the region $\RR_+^2$ is strongly invariant for
  the $2$-species system in the sense that $\rho_1\mn\rho_2$ becomes strictly positive (and sufficiently fast) for the solution
  $\bs{\rho}$, then one can replace the assumption $\bs{\rho}([0,T_{\rm{max}})\x\T^d)\subs(0,+\infty)^2$ with the assumption
    $\bs{\rho}_0\geq 0$ as in~\cite[Remark 3.3]{Stamatakis2015a} for the $1$-species case. Secondly, if one knew that the region
    $(0,+\infty)^2$ is invariant for the $2$-species ZRP system, starting from $C^{2+\theta}$ non-negative initial data
    $\bs{\rho}_0\colon\T^d\to\bs{R}(\DD_{\bs{R}}^o)$, one could could choose small enough $\ee>0$ such that
    $\bs{\rho}_0^\ee(\T^d)\subs\bs{R}(\DD_{\bs{R}}^o)\cap(0,+\infty)^2$ where $\rho^\ee_{0,i}=\rho_{0,i}+\ee$, $i=1,2$, use the
    result for strictly positive data and try to pass to the limit as $\ee\to 0$. Since we do not pursue the study of the
    $2$-species ZRP system and its invariant regions at the macroscopic level in this article, we consider only local solutions
    which are strictly positive and sub-critical and whose existence is established by Amann~\cite{Amann1986a}.

The next result states that, when starting from sufficiently regular subcritical initial profiles, the species-blind
system~\eqref{SumSystemPDE} has solutions defined globally in time.

\begin{theorem}[Regularity and global existence for the species-blind system]
  \label{PDE}
  Let $\bs{\rho}_0\in C^{2+\theta}(\T^d;\bs{R}(\DD_{\bs{R}}^o)\cap(0,\infty)^2)$, $\theta\in[0,1)$, be an initial profile. Then
    the species-blind parabolic system~\eqref{SumSystemPDE} has a unique classical solution $\bs{\rho}\colon\RR_+\x\T^d\to\RR^2$
    starting from $\bs{\rho}_0$ and
  \begin{equation*}
    \bs{\rho}\in C^{1+\theta,2+\theta}_{\rm{loc}}([0,+\infty)\x\T^d;\bs{R}(\DD_{\bs{R}}^o)\cap(0,\infty)^2).
  \end{equation*}
\end{theorem}
     	
The proof of this Theorem can be found in Subsection~\ref{sec:Proof-Theor} and it is obtained by taking into account the fact that
the sum $\rho_1+\rho_2$ of the two variables of a solution $\bs{\rho}=(\rho_1,\rho_2)$ of the $2$-species blind system is a
solution of the scalar parabolic equation $\pd_t(\rho)=\Delta\hat{\Phi}(\rho)$. Here, by using the strong maximum principle for
scalar quasilinear parabolic equations and by proving that classical solutions $\bs{\rho}$ of the species-blind parabolic system
do not become negative, we obtain that the the sub-critical region is an invariant region. We believe that $\RR_+^2$ will be an
invariant region of the $2$-species ZRP parabolic system in general. Yet, since we do not study this question in this article, in
order to be rigorous we prove it in this particular case. We should add that the arguments used strongly rely on the relation to
the PDE of the single species ZRP associated to the species-blind ZRP by ``ignoring'' the species, and thus do not easily extend
to the general case.

As a corollary, we find that the hydrodynamic limit for the species-blind process holds globally in time;
Subsection~\ref{sec:Proof-Coroll} gives the proof.

\begin{cor}
  \label{SpBlHL} 
  Let $(S_t^N)_{t\geq 0}$ be the transition semigroup of the diffusively rescaled species-blind symmetric n.n.~ZRP on the torus
  $\T_N^d$ corresponding to a $1$-species jump rate $\hat{g}$ such that
  $\hat{\phih}_c:=\liminf_{k\ra+\infty}\hat{g}!(k)^\fr{k}\in(0,+\infty]$ exists as a limit. Assume further that $\hat{g}$ is
    bounded if the critical density $\hat{\rho}_c$ of the $1$-species ZRP is finite. If $\mu_0^N\in\PP(\MM_N^{d;2})$ is an entropy
    local equilibrium of profile $\bs{\rho}_0\in C^{2+\theta}(\T^d;\bs{R}(\DD_{\bs{R}}^o)\cap(0,\infty)^2)$, then
    $\mu_t^N:=\mu_0^NS_t^N$ is an entropy local equilibrium of profile $\bs{\rho}(t,\cdot)$ for all $t\geq 0$, where $\bs{\rho}\in
    C^{1+\theta,2+\theta}_{\rm{loc}}(\RR_+\x\T^d;\bs{R}(\DD_{\bs{R}}^o)\cap(0,\infty)^2)$ is the unique solution to the
    species-blind parabolic system~\eqref{SumSystemPDE} starting from $\bs{\rho}_0$.
\end{cor}

\section{Proofs}
\label{Proofs}

\subsection{Proof of Theorem~\protect\ref{OBE}}
\label{sec:Proof-Prop}

The proof of the one-block estimate follows closely the proof for the one-species case found in~\cite[Section
  5.4]{Kipnis1999a}. The differences are twofold. In~\cite[Section 5.4]{Kipnis1999a}, the one-species case is treated, and we
extend this result to two species. However, the main difference is that in~\cite{Kipnis1999a} the non-condensing case is treated,
while we cover the condensing case as well. This is shown by applying the equivalence of ensembles~\eqref{EqOfEns} as
in~\cite{Stamatakis2015a}.

The first step in the proof of the one-block estimate is to replace the jump rate $\bs{g}(\bs{\eta}(x))$ at the site $x$ with the
spatial average $\bs{g}(\bs{\eta}(x))^\ell$ over a box of size $\ell\in\NNZ$. This is based on the following lemma which is also
useful in the proof of Theorem~\ref{HL}. The proof is omitted as it is a simple adaptation of the proof for the one-species
case~\cite[Lemma 6.4.1]{Kipnis1999a}.
\begin{lemma}
  \label{GlobPartDensEst} 
  If the sequence $\{\mu_0^N\}$ of initial distributions satisfies the $O(N^d)$-entropy assumption~\eqref{EntrAssumpt}, then 
  \begin{equation*}
    \int|\bs{\eta}|_1d\mu_0^N\leq O(N^d),
  \end{equation*}
  where $|\bs{\eta}|_1 := |\bs{\eta}|_{N,1}:=\sum_{x\in\T_N^d}|\bs{\eta}(x)|_1$.
\end{lemma}     		
     		
This lemma, a change of variables and the conservation of the number of particles allow us to replace $\bs{g}(\bs{\eta}(x))$ with
the spatial average $\bs{g}(\bs{\eta}(x))^\ell$ in the statement of the one-block estimate, and thus the one-block estimate is
reduced to proving that
\begin{align}
  \label{ellSpatTimeDependentOneBlockEstimate}
  \lim_{\ell\ra\infty}\limsup_{N\ra\infty}\int\fr{N^d}\sum_{x\in\T^d}\tau_xV^\ell d\bar{\mu}_T^N=0,
\end{align}
where $\bar{\mu}_T^N:=\fr{T}\int_0^T\mu_t^Ndt$ and $V^\ell$ is the cylinder function
$V^\ell:=|\bs{g}(\bs{\eta}(0))^\ell-\bar{\bs{\Phi}}(\bs{\eta}(0)^\ell)|_1$.

We establish this identity in a sequence of steps. We first estimate the entropy and the Dirichlet form of the density
$\bar{f}_T^N:={d\bar{\mu}_T^N}/{d\nu_{\bs{\rho}_*}^N}$ of $\bar{\mu}_T^N$ with respect to an equilibrium state of density
$\bs{\rho}_*\in A$. Note that $\bar{f}_T^N=\fr{T}\int f_t^Ndt$, where $f_t^N:={d\mu_t^N}/{d\nu_{\bs{\rho}_*}^N}$ is the density of
the law $\mu_t^N$ of the ZRP at time $t$ with respect to the product equilibrium state of density $\bs{\rho}_*\in A$.
By~\cite[Proposition A.9.1]{Kipnis1999a}, for any initial probability measure $\mu$ the entropy $H(\mu_t|\nu)$ of the law
$\mu_t:=\mu P_t$ of a Markov semigroup $(P_t)_{t\geq 0}$ at time $t$ with respect to an equilibrium state $\pi$ of $(P_t)$ is a
non-increasing function of time. Here the equilibrium $\pi$ need not be unique or approached by $\mu_t$ as
$t\ra+\infty$. Therefore, since $\mu_0^N$ satisfies the $O(N^d)$-entropy assumption, we have for fixed $\bs{\rho}_*\in A$ that
$H(\mu_t^N|\nu_{\bs{\rho}_*}^N)\leq C(\bs{\rho}_*)N^d$, which, by convexity of the entropy, implies that
$H(\bar{\mu}_T^N|\nu_{\bs{\rho}_*}^N)\leq C(\bs{\rho}_*)N^d$. Furthermore, if $D_N\colon L^1_+(\nu_{\bs{\rho}_*}^N)\to[0,+\infty]$
denotes the functional defined by $D_N(f)=\mathfrak{D}_N(\sqrt{f})$ where
$\mathfrak{D}_N\colon L^2(\nu_{\bs{\rho}_*})\to[0,+\infty]$ is the Dirichlet form associated to the generator $L_N$,
\begin{equation*}
  \mathfrak{D}_N(f):=-\ls f,L_Nf\rs_{\nu_{\bs{\rho}_*}}=-\int fL_Nfd\nu_{\bs{\rho}_*},
\end{equation*} 
then by~\cite[Proposition A.9.2]{Kipnis1999a} and the convexity of the functional $D_N$, it follows that
$D_N(\bar{f}_T^N)\leq\fr{T}\int_0^TD_N(f_t^N)dt\leq\frac{C(\bs{\rho}_*)}{2T}N^{d-2}$. Therefore, if we set
$H_N(f):=H(fd\nu_{\bs{\rho}_*}^N|\nu_{\bs{\rho}_*}^N)$, in order to prove the one-block estimate, it suffices to prove that for
some $\bs{\rho}_*\in A$
\begin{equation}
  \label{StaticOBE}
  \limsup_{\ell\ra\infty}\limsup_{N\ra\infty}\sup_{\substack{H_N(f)\leq C_0N^d\\D_N(f)\leq 
      C_0N^{d-2}}}\int\fr{N^d}\sum_{x\in\T_N^d}\tau_xV^\ell fd\nu_{\bs{\rho}_*}^N\leq 0,\quad\forall\;C_0>0,
\end{equation}
where the supremum is taken among all densities $f\in L^1_+(\nu_{\bs{\rho}_*}^N)$.
     		
In a second step, following the proof of the one-species case~\cite[Section 5.4]{Kipnis1999a} we cut off large densities. Since
Lemma~\ref{GlobPartDensEst} requires only the $O(N^d)$-entropy assumption, it follows that
\begin{align}
  \label{Basic}
  \limsup_{N\ra+\infty}\sup_{H_N(f)\leq CN^d}\fr{N^d}\int|\bs{\eta}|_1fd\nu_{\bs{\rho}_*}^N<+\infty, \text{ for every } C>0.
\end{align}
Similarly to the $1$-species case, under the assumption that $\bs{g}$ has sublinear growth at infinity in the sense
of~\eqref{SubLinGrowth} (which always holds when $\bs{g}$ is bounded), inequality~\eqref{Basic} allows us to cut off large
densities, by restricting $V^\ell$ to the set of configurations $\bs{\eta}$ which satisfy $|\bs{\eta}^\ell(0)|_1\leq C_1$ for some
constant $C_1>0$.
This way the one-block estimate is reduced to proving that for all constants $C_0,C_1>0$
\begin{equation}
  \label{StaticFinal}
  \lim_{\ell\ra+\infty}\limsup_{N\ra+\infty}\sup_{D_N(f)\leq C_0N^{d-2}}\int\fr{N^d}\sum_{x\in\T_N^d}\tau_xV^\ell
  \1_{\{|\bs{\eta}^\ell(x)|_1\leq C_1\}}fd\nu_{\bs{\rho}_*}^N\leq 0.
\end{equation} 
In a third step, by adapting the steps 2 to 4 of~\cite[Sect. 5.4.1]{Kipnis1999a} to the two-species case, the one-block-estimate
is further reduced to showing that for all constants $C_1>0$,
\begin{align}
  \label{step5end}
  \limsup_{\ell\ra+\infty}\max_{\bs{K} \bigm| |\bs{K}|_1\leq(2\ell+1)^dC_1}\int V^\ell d\nu_{2\ell+1,\bs{K}}=0,
\end{align}
where the canonical measure $\nu_{2\ell+1,\bs{K}}$ is considered as a measure on $\MM_\infty^d$ by identifying the cube
$\Lambda_\ell^d:=\{x\in\ZZ^d\bigm||x|\leq\ell\}\subs\ZZ^d$ with $\T_{2\ell+1}^d$.

The final step in the proof of the one-block estimate consists in applying the equivalence of ensembles to
prove~\eqref{step5end}. Since the measure $\nu_{2\ell+1,\bs{K}}$ is concentrated on configurations with $\bs{K}$ particles, the
integral appearing in~\eqref{step5end} is equal to
\begin{equation*}
  \int V^\ell d\nu_{2\ell+1,\bs{K}}=\int\bigg|\fr{(2\ell+1)^d}\sum_{|x|\leq\ell}\bs{g}\big(\bs{\xi}(x)\big)
  -\bar{\bs{\Phi}}\Big(\frac{\bs{K}}{(2\ell+1)^d}\Big)\bigg|_1d\nu_{2\ell+1,\bs{K}}.
\end{equation*}
As in the one-species case, by fixing a positive integer $k$ which will tend to infinity after taking the limit as
$\ell\ra+\infty$, and decomposing the cube $\Lambda_\ell^d$ in smaller cubes of side-length $2k+1$, the one-block estimate is
reduced to showing that
\begin{align}
  \label{FinalLimInOBE}	
  \lim_{k\ra\infty}\lim_{m\ra\infty}S(m,k)=0,
\end{align}
where $S(m,k)$ denotes the supremum
\begin{equation*}S(m,k):=\sup_{\substack{\ell\geq m\\|\bs{K}|_1\leq(2\ell+1)^dC_1}}
  \int\Big|\fr{(2k+1)^d}\sum_{|x|\leq k}\bs{g}\big(\bs{\xi}(x)\big)-
  \bar{\bs{\Phi}}\Big(\frac{\bs{K}}{(2\ell+1)^d}\Big)\Big|_1d\nu_{2\ell+1,\bs{K}}.
\end{equation*} 
This is the part of the proof where we need the boundedness and the regularity of the tails~\eqref{RegTails} of the jump rate
$\bs{g}$ as well as the continuity of the partition function $Z$ on $\DD_Z\cap\pd\DD_Z$. For each fixed $(m,k)\in\NN\x\NN$, we
pick a sequence $\{(\ell_n^{m,k},\bs{K}^{m,k}_n)\}_{n\in\NN}$ such that $\ell_n^{m,k}\geq m$ and
$|\bs{K}_n^{m,k}|_1\leq(2\ell_n^{m,k}+1)^dC_1$ for all $n\in\NN$ that achieves the supremum, i.e., such that
\begin{equation*}	
  S(m,k)=\lim_{n\ra\infty}\int\Big|\fr{(2k+1)^d}\sum_{|x|\leq k}\bs{g}\big(\bs{\xi}(x)\big)-
  \bar{\bs{\Phi}}\Big(\frac{\bs{K}^{m,k}_n}{(2\ell_n^{m,k}+1)^d}\Big)\Big|_1d\nu_{2\ell_n^{m,k}+1,\bs{K}^{m,k}_n}.
\end{equation*}
Since the sequence $\{\bs{r}_n^{m,k}\}_{n\in\NN}$ defined by
\begin{equation*}
  \bs{r}_n^{m,k}:=\frac{\bs{K}^{m,k}_n}{(2\ell_n^{m,k}+1)^d},\qquad n\in\NN,
\end{equation*} 
is contained in the compact triangular region $B_{|\cdot|_1}(0,C_1):=\{\bs{r}\in\RR_+^2 \bigm| |\bs{r}|_1\leq C_1\}$, for each
fixed $(m,k)\in\NN\x\NN$, we can pick a sequence $\{n_j\}_{j\in\NN} := \{n_j^{m,k}\}$ such that $\bs{r}^{m,k}_{n_j}$ converges to
some $\bs{r}^{m,k}\in B_{|\cdot|_1}(0,C_1)$ as $j\ra\infty$.  Since we assume that $\bs{g}$ is bounded, it follows by the
equivalence of ensembles that
\begin{equation*}
  S(m,k)=\int\Big|\fr{(2k+1)^d}\sum_{|x|\leq k}\bs{g}\big(\bs{\xi}(x)\big)-
  \bar{\bs{\Phi}}\big(\bs{r}^{m,k}\big)\Big|_1d\nu_{\bs{R}_c(\bs{r}^{m,k})}^\infty.
\end{equation*}
Furthermore, since $|\bs{R}_c(\bs{\rho})|_1\leq|\bs{\rho}|_1$, for each fixed $k\in\NN$ the sequence
$\{\bs{\rho}^{m,k}:=\bs{R}_c(\bs{r}^{m,k})\}_{m\in\NN}$, is also contained in $B_{|\cdot|_1}(0,C_1)$ and thus we can choose a
sequence $\{m_j\}_{j\in\NN}=\{m_j^{(k)}\}$ such that $\{\bs{\rho}^{m_j,k}\}_{m\in\NN}$ converges to some
$\bs{\rho}^k\in B_{|\cdot|_1}(0,C_1)\cap\bs{R}(\DD_{\bs{R}})$. By the continuity assumption on $Z$, the grand canonical ensemble
is weakly continuous. By this fact, the continuity of $\bs{R}_c$ and the identity $\bar{\bs{\Phi}}=\bs{\Phi}\circ\bs{R}_c$,
\begin{align*}
  \lim_{m\ra\infty}S(m,k)=\int\Big|\fr{(2k+1)^d}\sum_{|x|\leq k}\bs{g}\big(\bs{\xi}(x)\big)-
  \bs{\Phi}\big(\bs{\rho}^k\big)\Big|_1d\nu_{\bs{\rho}^k}^\infty.
\end{align*}
Therefore
\begin{equation*}
  \limsup_{k\ra+\infty}\lim_{m\ra+\infty}S(m,k)\leq
  \limsup_{k\ra\infty}\sup_{\bs{\rho}\in\bs{R}(\DD_{\bs{R}})}\int\Big|\fr{(2k+1)^d}
  \sum_{|x|\leq k}\bs{g}\big(\bs{\eta}(x)\big)-
  \bs{\Phi}\big(\bs{\rho}\big)\Big|_1d\nu_{\bs{\rho}}^\infty.
\end{equation*}
The random variables $\bs{g}\big(\bs{\eta}(x)\big)$, $x\in\ZZ^d$, are uniformly bounded by $\|\bs{g}\|_\infty$ and i.i.d.~with
respect to $\nu_{\bs{\rho}}^\infty$ for all $\bs{\rho}\in\bs{R}(\DD_{\bs{R}})$ and thus they satisfy the $L^2$-weak law of large
numbers uniformly over all parameters $\bs{\rho}\in\bs{R}(\DD_{\bs{R}})$, which shows that the term in the right hand side above
is equal to zero. This completes the proof of the one-block estimate and hence the proof of Theorem~\ref{OBE}. \qed

\subsection{Proof of Theorem~\protect{\ref{HL}}}
\label{sec:Proof-Theor-1}

Let $A$ be the interior of the set of all strictly positive sub-critical densities, i.e.,
\begin{equation*}
  A:=\bs{R}(\DD_{\bs{R}}^o)\cap(0,\infty)^2,
\end{equation*} 
and let $\bs{\rho}\colon[0,T_{\rm{max}})\x\T^d\to A$ be the maximal classical solution established in~\cite{Amann1986a} of the
initial value problem~\eqref{NonLinDiffSystem} with $\bs{\rho}(0,\cdot) := \bs{\rho}_0\in C^{2+\theta}(\T^d;A)$. We fix
$\bs{a}\in A$ and denote by $\psi^{N}_t$ the Radon-Nikodym derivative of $\nu_{\bs{\rho}_t(\cdot)}^N$ with respect to
$\nu_{\bs{a}}^N$,
\begin{equation*}
  \psi_t^N:=\frac{d\nu_{\bs{\rho}_t(\cdot)}^N}{d\nu_{\bs{a}}^N}.
\end{equation*} 
Let $H_N(t):=\HHH(\mu_t^N|\nu_{\bs{\rho}_t(\cdot)}^N)$ be the relative entropy of $\mu_t^N$ with respect to
$\nu_{\bs{\rho}_t(\cdot)}^N$. We have the following upper bound on the entropy production, proved in~\cite[Lemma
6.1.4]{Kipnis1999a},
\begin{equation}
  \label{AFirstUpBoundOnEntrProd}
  \pd_tH_N(t)\leq\int\fr{\psi_t^N}\big\{N^2L_N^*\psi_t^N-\pd_t\psi_t^N\big\}d\mu_t^N
\end{equation}
for every $t\in[0,T_{\rm{max}})$, where $L_N^*$ is the adjoint of $L_N$ in $L^2(\nu_{\bs{a}}^N)$. Denoting by 
\begin{equation}
  \label{LRE}
  H(t):=\limsup_{N\ra\infty}\fr{N^d}H_N(t),\quad t\in[0,T_{\rm{max}}),
\end{equation} 
the limiting entropy density, the main step in the application of the relative entropy method is to use this upper bound on
$\pd_tH_N(t)$ to get an inequality of the form
\begin{equation}
  \label{GronwIneqOnLimitingRenormedEntropy}
  H(t)\leq H(0)+ \fr{\g}\int_0^tH(s)ds
\end{equation} 
for some constant $\g>0$. Since $H(0)=0$ by assumption, this implies by Gronwall's inequality that $H(t)=0$ for all
$t\in[0,T_{\rm{max}})$ as required. Of course, in order for Gronwall's inequality to be applicable, $H$ must belong at least in
$L^1_{{\rm{loc}}}([0,T_{\rm{max}}))$. This is the context of the next two lemmas. The first is Remark 6.1.2 in~\cite{Kipnis1999a}
for single-species ZRPs.

\begin{lemma}
  \label{lemmaLitlleoToBigOEntrAssumpt} 
  If $\{\mu_0^N\}$ is an entropy local equilibrium of profile $\bs{\rho}\in C(\T^d;\bs{R}(\DD_{\bs{R}}^o))$, then $\{\mu_0^N\}$
  satisfies the $O(N^d)$-entropy assumption~\eqref{EntrAssumpt}.
\end{lemma}

\proof Indeed, for fixed $\bs{a}\in A:=\bs{R}(\DD_{\bs{R}}^o)\cap(0,\infty)^2$, by the relative entropy
inequality~\cite[Section A.1.8]{Kipnis1999a}
\begin{equation}
  \label{lemma:LitllFirstBound}
  \HHH(\mu_0^N|\nu_{\bs{a}}^N)
  \leq\Big(1+\fr{\g}\Big)\HHH(\mu_0^N|\nu_{\bs{\rho}(\cdot)}^N)+\fr{\g}\log\int e^{\g\log\frac{d\nu_{\bs{\rho}(\cdot)}^N}{d\nu_{\bs{a}}^N}}d\nu_{\bs{\rho}(\cdot)}^N.
\end{equation}
Since $\nu_{\bs{\rho}(\cdot)}^N$, $\nu_{\bs{a}}^N$ are product measures, the Radon-Nikodym derivative
$\frac{d\nu_{\bs{\rho}(\cdot)}^N}{d\nu_{\bs{a}}^N}$ can be computed explicitly.
With the notation $\bs{\Phi}_{\bs{a}}:=\big(\frac{\Phi_1}{\Phi_1(\bs{a})},\frac{\Phi_2}{\Phi_2(\bs{a})}\big)$,
$Z_{\bs{a}}:=\frac{Z\circ\bs{\Phi}}{Z(\bs{\Phi}(\bs{a}))}$
\begin{equation*}
  \int \Big(\frac{d\nu_{\bs{\rho}(\cdot)}^N}{d\nu_{\bs{a}}^N}\Big)^\g 
  d\nu_{\bs{\rho}(\cdot)}^N=\prod_{x\in\T_N^d}\fr{Z_{\bs{a}}(\bs{\rho}(x/N))^\g}\int 
  e^{\ls\bs{k},\g\log\bs{\Phi}_{\bs{a}}(\bs{\rho}(x/N))\rs}d\nu_{\bs{\rho}(x/N)}^1(\bs{k}).
\end{equation*}
Since $Z\geq 1$, we have that
$\fr{Z_{\bs{a}}(\bs{\rho})}=\frac{Z(\bs{\Phi}(\bs{a}))}{Z({\bs{\Phi}}(\bs{\rho}))}\leq Z(\bs{\Phi}(\bs{a}))$ and therefore
\begin{align}
  \label{lemma:LitllSecBound}
  \fr{\g N^d}\log\int \Big(\frac{d\nu_{\bs{\rho}(\cdot)}^N}{d\nu_{\bs{a}}^N}\Big)^\g d\nu_{\bs{\rho}(\cdot)}^N&\leq 
  Z(\bs{\Phi}(\bs{a}))+\fr{\g N^d}\sum_{x\in\T_N^d}\Lambda_{\bs{\rho}(x/N)}
  \big(\g\log\bs{\Phi}_{\bs{a}}\big(\bs{\rho}\big(x/N\big)\big)\big)\nonumber\\
  &\leq Z(\bs{\Phi}(\bs{a}))+\fr{\g N^d}\sum_{x\in\T_N^d}\log Z\big(\bs{F}_{\bs{a}}(x/N,\g)\big),
\end{align}
where here $\bs{F}_{\bs{a}}\colon\T^d\x[0,1]\to(0,\infty)^2$ is the function given by
$\bs{F}_{\bs{a}}(u,\g)=\frac{\bs{\Phi}(\bs{\rho}(u))^{1+\g}}{\bs{\Phi}(\bs{a})^\g}$ and for $\bs{a}\in\RR_+^2$,
$\bs{b}\in(0,\infty)^2$, $\g>0$, we have set $\bs{a}^\g:=(a_1^\g,a_2^\g)$ and
$\frac{\bs{a}}{\bs{b}}:=(\frac{a_1}{b_1},\frac{a_2}{b_2})$. Since $\bs{\rho}(\T^d)\subs\bs{R}(\DD_{\bs{R}}^o)$ by assumption, it
follows that $\bs{\Phi}(\bs{\rho}(\T^d))\subs\DD_Z^o$. Since $\bs{F}_{\bs{a}}$ is uniformly continuous on $\T^d\x[0,1]$ and
satisfies $\lim_{\g\ra 0}\bs{F}_{\bs{a}}(u,\g)=\bs{\Phi}(\bs{\rho}(u))$ for all $u\in\T^d$, it follows that its image is contained
in $\DD_Z^o$, i.e., $\{\bs{F}_{\bs{a}}(u,\g)|u\in\T^d\}\subs\DD_Z^o$ for sufficiently small $\g>0$. Then the function $u\mapsto
Z(\bs{F}_{\bs{a}}(u,\g))$ is well defined and continuous on the torus $\T^d$, so that its Riemannian sums
converge. By~\eqref{lemma:LitllFirstBound}, \eqref{lemma:LitllSecBound} and the fact that $\mu_0^N$ is an entropy local
equilibrium, this yields that
\begin{equation*}
  C(\bs{a})\leq Z(\bs{\Phi}(\bs{a}))+\fr{\g}\int_{\T^d}\log Z\big(\bs{F}_{\bs{a}}(u,\g))du<+\infty
\end{equation*} 
for small $\g>0$, and the proof of Lemma~\ref{lemmaLitlleoToBigOEntrAssumpt} is complete. \qed

\begin{lemma}
  \label{lemmaUnifBoundNormalizedMicroscopicEntropies}
  Let $\bs{\rho}\colon[0,T]\x\T^d\to \bs{R}(\DD_{\bs{R}}^o)\cap(0,\infty)^2$ be a continuous function and let $\{\mu_0^N\}$ be an
  entropy local equilibrium with respect to $\bs{\rho}_0:=\bs{\rho}(0,\cdot)$. Then the upper entropy $\bbar{H} \colon
  [0,T]\to[0,+\infty]$ defined by
  \begin{equation*}
    \bbar{H}(t):=\sup_{N\in\NN}\fr{N^d}\HHH(\mu_t^N|\nu_{\bs{\rho}_t(\cdot)}^N)
  \end{equation*} 
  belongs to $L^\infty([0,T])$.
\end{lemma}

\proof By the relative entropy inequality and~\cite[Proposition A.1.9.1]{Kipnis1999a}, according to which the function
$t\mapsto \HHH(\mu_t^N|\nu_{\bs{a}}^N)$ is non-increasing,
\begin{align}
  \label{REBound2}
  H_N(t)\leq\Big(1+\fr{\g}\Big)\HHH(\mu_0^N|\nu_{\bs{a}}^N)+\fr{\g}\log
  \int\Big(\frac{d\nu_{\bs{a}}^N}{d\nu_{\bs{\rho}_t(\cdot)}^N}\Big)^\g d\nu_{\bs{a}}^N
\end{align} 
for all $t\geq 0$ and all $\g>0$. Since the proper domain of $Z_{\bs{a}}$ has interior $\DD_{Z_{\bs{a}}}^o=\bs{R}(\DD_{\bs{R}}^o)$
and since $\bs{\rho}([0,T]\x\T^d)\subs\bs{R}(\DD_{\bs{R}}^o)\cap(0,\infty)^2$, the function $Z_{\bs{a}}\circ\bs{\rho}$ is a
bounded continuous function on the torus, and therefore, by a computation similar to the one in the proof of
Lemma~\ref{lemmaLitlleoToBigOEntrAssumpt}, we obtain
\begin{equation*}
  \fr{\g N^d}\log\int\Big(\frac{d\nu_{\bs{a}}^N}{d\nu_{\bs{\rho}_t(\cdot)}^N}\Big)^\g d\nu_{\bs{a}}^N
  =\|Z_{\bs{a}}\circ\bs{\rho}\|_{L^\infty([0,T]\x\T^d)}
  +\fr{\g N^d}\sum_{x\in\T_N^d}\Lambda_{\bs{a}}\Big(\g\log\fr{\bs{\Phi}_{\bs{a}}(\bs{\rho}_t(x/N))}\Big),
\end{equation*}
where $\|Z_{\bs{a}}\circ\bs{\rho}\|_{L^\infty([0,T]\x\T^d)}<+\infty$. For the second term, we have for every $u\in\T^d$ that
\begin{align}
  \label{LogMom}
  \Lambda_{\bs{a}}\Big(\g\log\fr{\bs{\Phi}_{\bs{a}}(\bs{\rho}_t(u))}\Big)=\log\Big\{\fr{Z(\bs{\Phi}(\bs{a}))}
  Z\Big(\frac{\bs{\Phi}(\bs{a})^{1+\g}}{\bs{\Phi}(\bs{\rho}_t(u))^{\g}}\Big)\Big\}.
\end{align} 
Since $\bs{\Phi}(\bs{\rho})(\T^d)\subs(0,\infty)^2$ and $\bs{\Phi}(\bs{\rho})$ is continuous, there exists $\bs{\phih}_0\in\DD_Z$
such that $\bs{\phih}_0<\bs{\Phi}(\bs{\rho}(t,u))$ for all $(t,u)\in[0,T]\x\T^d$. Then since $Z$ is increasing,
\begin{equation*} 
  Z\Big(\frac{\bs{\Phi}(\bs{a})^{1+\g}}{\bs{\Phi}(\bs{\rho}_t(u))^{\g}}\Big)\leq  
  Z\Big(\frac{\bs{\Phi}(\bs{a})^{1+\g}}{\bs{\phih}_0^\g}\Big),
\end{equation*} 
and since $\bs{\Phi}(\bs{a})\in\DD_Z^o$ and $\bs{\Phi}(\bs{a})^{1+\g}/\bs{\phih}_0^\g\to\bs{\Phi}(\bs{a})$ as $\g\ra 0$, we can
choose $\g_0>0$ sufficiently small so that $\bs{\Phi}(\bs{a})^{1+\g}/\bs{\phih}_0^\g\in\DD_Z^o$ and
$Z\big({\bs{\Phi}(\bs{a})^{1+\g}}/{\bs{\phih}_0^\g}\big)\leq Z\big(\bs{\Phi}(\bs{a})\big)+1$ for all $\g<\g_0$. Consequently,
since by Lemma~\ref{lemmaLitlleoToBigOEntrAssumpt} $\{\mu_0^N\}$ satisfies the $O(N^d)$-entropy assumption, by~\eqref{REBound2}
for some constant $C\geq 0$ for all $\g<\g_0$
\begin{equation*}  
  \|\bbar{H}\|_{L^\infty([0,T])}\leq\Big(1+\fr{\g}\Big)C+\|Z_{\bs{a}}(\bs{\rho})\|_{L^\infty([0,T]\x\T^d)}+\fr{\g}
  \log\frac{Z(\bs{\Phi}(\bs{a}))+1}{Z(\bs{\Phi}(\bs{a}))}<+\infty,
\end{equation*}
establishing the claim of Lemma~\ref{lemmaUnifBoundNormalizedMicroscopicEntropies}.
\qed
     		
The bound~\eqref{AFirstUpBoundOnEntrProd} on the entropy production can be estimated explicitly. Since
$\nu_{\bs{\rho}_t(\cdot)}^N$, $\nu_{\bs{a}}^N$ are product measures, $\psi_t$ can be computed explicitly. Then by differentiating,
using the chain rule, the fact that $\bs{\rho}$ is a solution of the hydrodynamic equation, the relations
$\frac{\phih_i\pd_iZ(\bs{\phih})}{Z(\bs{\phih})}=R_i(\bs{\phih})$, $i=1,2$ and the relation~\eqref{MacrCompCond} we obtain
%
\begin{align}
  \label{Expl1}
  \frac{\pd_t\psi_t^N}{\psi_t^N}=\sum_{x\in\T_N^d} \Big\ls\frac{\Delta\bs{\Phi}(\bs{\rho}_t(x/N))}{\bs{\Phi}(\bs{\rho}_t(x/N))},
  D\bs{\Phi}\big(\bs{\rho}_t(x/N)\big)[\bs{\eta}(x)-\bs{\rho}_t(x/N)]\Big\rs.
\end{align}
As already noted in~\cite{Grosskinsky2003b}, in this computation in the two-species case, one has to use the macroscopic
analogue~\eqref{MacrCompCond} of the compatibility relations~\eqref{CompatibilityCondition}.

For the other term, by computations of the action of the generator on $\psi_t^N$ similar to the ones for the single-species case
in~\cite{Kipnis1999a},
\begin{align}
  \label{ForTaylor}  
  \frac{L_N^*\psi_t^N}{\psi_t^N}=\sum_{i=1,2}\sum_{x,y\in\T_N^d}\Big[\frac{\Phi_i\big(\bs{\rho}_t(y/N)\big)}
  {\Phi_i\big(\bs{\rho}_t(x/N)\big)}-1\Big] 
  \big[g_i\big(\bs{\eta}(x)\big)-\Phi_i\big(\bs{\rho}_t(x/N)\big)\big]p(y-x).
\end{align}
Since $\bs{\Phi}(\bs{\rho}_t)$ is $C^{2+\theta}$ for some $\theta>0$ and the n.n.~transition probability has mean zero, the Taylor
expansion for $C^{2+\theta}$ functions yields (with the renormalisation $p(\ZZ^d)=2d$) that
\begin{align}
  \label{Expl2}
  \frac{N^2L_N^*\psi_t^N}{\psi_t^N}=\sum_{x\in\T_N^d}
  \Big\ls\frac{\D[\bs{\Phi}(\bs{\rho}_t)]}{\bs{\Phi}(\bs{\rho}_t)}\big(\frac{x}{N}\big),\bs{g}(\bs{\eta}(x))
  -\bs{\Phi}(\bs{\rho}_t(x/N))\Big\rs+r_N(t).
\end{align}
Here, for any $T\in[0,T_{\rm{max}})$, the remainder $r_N(t)$ satisfies the bound
\begin{equation*}
  |r_N(t)|\leq\frac{C_T\bs{g}^*}{N^\theta}|\bs{\eta}|_1+\frac{C_TM_T}{m_T}N^{d-\theta}
\end{equation*}
for all $t\in[0,T]$, where $\bs{g}^*$ is the constant in~\eqref{quasiLipschitz1}, $C_T=C(d,p,\bs{\Phi}(\bs{\rho}),T)\geq 0$ is the
constant
\begin{equation*}
  C_T=\sqrt{d}\sup_{0\leq t\leq T}\|D^2[\Phi_1(\bs{\rho}_t)]\|_{C^{\theta}}\mx
  \|D^2[\Phi_2(\bs{\rho}_t)]\|_{C^{\theta}}\sum_{y\in\ZZ^d}\|y\|^{2+\theta}p(y)
\end{equation*}
with $\|\cdot\|_{C^\theta}$ denoting the $\theta$-H\"{o}lder seminorm and
\begin{equation*}
  m_T:=\inf_{(t,u)\in[0,T]\x\T^d}\min_{i=1,2}\Phi_i(\bs{\rho}_t(u))>0,\qquad
  M_T:=\sup_{(t,u)\in[0,T]\x\T^d}|\bs{\Phi}(\bs{\rho}(t,u))|_1<+\infty.
\end{equation*} 
By this bound on the remainder and the conservation of the number of particles it follows that for all
$t\in[0,T]\subs[0,T_{\rm{max}})$, $T>0$,
\begin{equation*}
  \fr{N^d}\int_0^t\int r_N(t)d\mu_t^Ndt\leq \frac{C_T\bs{g}^*t}{N^{d+\theta}}\int|\bs{\eta}|_1d\mu_0^N+\frac{C_TM_Tt}{m_T}\fr{N^\theta},
\end{equation*}
which according to Lemma~\ref{GlobPartDensEst} shows that
\begin{align}
  \label{RemainderVanish}
  \int_0^t\int r_N(s)d\mu_s^Nds\leq o(N^d).
\end{align}
Since the function $\frac{\D[\bs{\Phi}(\bs{\rho}_t)]}{\bs{\Phi}(\bs{\rho}_t)}$ is in $C_{\rm{loc}}([0,T_{\rm{max}})\x\T^d)$, a change of
  variables shows that
\begin{align}
  \label{CV}
  \int_0^t\int\sum_{x\in\T_N^d}\Big\ls\frac{\D[\bs{\Phi}(\bs{\rho}_s)]}{\bs{\Phi}(\bs{\rho}_s)}
  \big(\frac{x}{N}\big),\bs{\eta}(x)-\bs{\eta}^\ell(x)\Big\rs d\mu_s^N ds=o(N^d)
\end{align}
for all $t\in[0,T_{\rm{max}})$. Integrating~\eqref{AFirstUpBoundOnEntrProd} in time, using the explicit expressions~\eqref{Expl1},
  \eqref{Expl2}, taking into account~\eqref{RemainderVanish} and the fact that $\{\mu_0^N\}$ is an entropy local equilibrium
  (i.e.,~\eqref{LittleEntrAss} holds) and using~\eqref{CV} and the one-block estimate, one obtains that for all
  $t\in(0,T_{\rm{max}})$
\begin{align}
  \label{AfterONEBLOCKApply}
  H_N(t)\leq
  \int_0^t\int\sum_{x\in\T_N^d}
  \Big\ls\frac{\D[\bs{\Phi}(\bs{\rho}_s)]}{\bs{\Phi}(\bs{\rho}_s)}
  \big(\frac{x}{N}\big),\bs{\Psi}\big(\bs{\rho}_s(x/N),\bs{\eta}^\ell(x)\big)\Big\rs d\mu_s^Nds+o_\ell(N^d),
\end{align} 
where $\bs{\Psi}\colon\bs{R}(\DD_{\bs{R}}^o)\x\RR_+^2\to\RR^2$ is the \emph{quasi-potential}
\begin{equation}
  \label{psi}
  \bs{\Psi}(\bs{\rho},\bs{\lambda})=\bar{\bs{\Phi}}(\bs{\lambda})-\bs{\Phi}(\bs{\rho})-D\bs{\Phi}(\bs{\rho})(\bs{\lambda}-\bs{\rho})
\end{equation} 
and the term $o_\ell(N^d)$ satisfies $o_\ell(N^d)/N^d\to 0$ as $N$ and then $\ell$ tend to infinity. In the definition of the
quasi-potential the second variable $\bs{\lambda}$ is in $\RR_+^2$ since it is to be substituted by the large microscopic averages
$\bs{\eta}^\ell(x)$, $x\in\T_N^d$. Thus the extension $\bar{\bs{\Phi}}$ of $\bs{\Phi}$ must be used in the quasi-potential. To
simplify the notation, we set
\begin{equation}
  \label{G}
  G_t(u,\bs{\lambda}):=\big\ls\frac{\D[\bs{\Phi}(\bs{\rho}_t)]}{\bs{\Phi}(\bs{\rho}_t)}(u),
  \bs{\Psi}\big(\bs{\rho}_t(u),\bs{\lambda}\big)\big\rs.
\end{equation} 
By the relative entropy inequality, we have for all $\g>0$ and all $0\leq s<T_{\rm{max}}$ that
\begin{equation*}
  \int\sum_{x\in\T_N^d}G_s\big(x/N,\bs{\eta}^\ell(x)\big)d\mu_s^N\leq\fr{\g}H_N(s)+\fr{\g}\log\int 
  e^{\g\sum_{x\in\T_N^d}G_s(\frac{x}{N},\bs{\eta}^\ell(x))}d\nu_{\bs{\rho}_s(\cdot)}^N.
\end{equation*}
By combining this inequality with the bound~\eqref{AfterONEBLOCKApply}, dividing by $N^d$ and taking the $\limsup$ as $N\ra\infty$
and then $\ell\ra\infty$, we get
\begin{equation}
  \label{AlmostGronwall}
  H(t)\leq\fr{\g}\int_0^tH(s)ds+\limsup_{\ell,N\ra+\infty}\fr{\g N^d}\int_0^t\log\int 
  e^{\g\sum_{x\in\T_N^d}G_s(\frac{x}{N},\bs{\eta}^\ell(x))}d\nu_{\bs{\rho}_s(\cdot)}^Nds,
\end{equation}
where in order to obtain the term $\int_0^tH(s)ds$ we used Lemma~\ref{lemmaUnifBoundNormalizedMicroscopicEntropies} to pass the
limit inside the integral and $\limsup_{\ell,N\ra+\infty}$ denotes the $\limsup$ as $N\ra+\infty$ and then $\ell\ra+\infty$.

To complete the proof of Theorem~\ref{HL}, it remains to show that for each $t\in[0,T_{\rm{max}})$ we can choose $\g>0$ small
  enough so that the rightmost term in~\eqref{AlmostGronwall} vanishes. We begin by noting that the function
  $G\colon[0,T_{\rm{max}})\x\T^d\x\RR_+^2\to\RR$ defined in~\eqref{G} satisfies
\begin{equation*}
  |G_t(u,\bs{\lambda})|     			
  \leq\Big|\frac{\D[\bs{\Phi}(\bs{\rho}_t)]}{\bs{\Phi}(\bs{\rho}_t)}(u)\Big|_\infty\Big\{\bs{g}^*\big(|\bs{\lambda}|_1
  +|(\bs{\rho}_t(u))|_1\big)+|D\bs{\Phi}(\bs{\rho}_t(u))|_\infty
  \big(|\bs{\lambda}|_1+|\bs{\rho}_t(u)|_1\big)\Big\}
\end{equation*}
for all $(t,u,\bs{\lambda})\in[0,T_{\rm{max}})\x\T^d\x\RR_+^2$, which for any $t\in[0,T_{\rm{max}})$ yields the inequality
\begin{equation}
  \label{UnifBoundOnGOnFinTImeHorizon}
  \sup_{(s,u)\in[0,t]\x\T^d}|G_s(u,\bs{\lambda})|\leq C_t\cdot(1+|\bs{\lambda}|_1) \quad\text{for all }\bs{\lambda}\in\RR_+^2
\end{equation} 
for some constant $C_t<+\infty$. Since for any $t\in[0,T_{\rm{max}})$ we have $\bs{\rho}([0,t]\x\T^d)\subs\bs{R}(\DD_{\bs{R}}^o)$
  the set $\bs{\Phi}(\bs{\rho})([0,T]\x\T^d)$ is bounded away from the critical densities $\bs{\phih}_c\in\pd\DD_Z$, and thus
  there exists $\ee>0$ such that
\begin{equation*}
  \sup_{(t,u)\in[0,T]\x\T^d}\Lambda_{\bs{\rho}(t,u)}(\bs{\lambda})<+\infty,\quad\forall\bs{\lambda}\in D(0,\ee),
\end{equation*}
i.e.,
$0\in\big(\DD_{\sup_{(t,u)\in[0,T]\x\T^d}\Lambda_{\bs{\rho}(t,u)}}\big)^o=\big(\bigcap_{(t,u)\in[0,T]\x\T^d}\DD_{\Lambda_{\bs{\rho}(t,u)}}\big)^o$. It
follows that by choosing $\g_t$ small enough so that $\g_tC_t<\ee$, we can pass the limit superior as $N\ra+\infty$ and then
$\ell\ra+\infty$ inside the time integral in the rightmost term in~\eqref{AlmostGronwall}. Thus in order to complete the proof it
remains to show that for each $t\in[0,T_{\rm{max}})$ we can choose $\g_t>0$ small enough so that
\begin{align}
  \label{AlmostGronwallTerm}
  \limsup_{\ell,N\ra+\infty}\fr{\g_t N^d}\log\int 
  e^{\g_t\sum_{x\in\T_N^d}G_t(\frac{x}{N},\bs{\eta}^\ell(x))}d\nu_{\bs{\rho}_t(\cdot)}^N\leq 0.
\end{align}

The proof of~\eqref{AlmostGronwallTerm} relies on a corollary of the Laplace-Varadhan lemma~\cite[Section 4.3]{Dembo2010a} for the
large deviations principle satisfied by the independent family of the occupation variables $\{\bs{\eta}(x)\}_{x\in\ZZ^d}$ with
respect to the invariant measure $\nu_{\bs{\rho}}^\infty$ on the infinite lattice $\ZZ^d$ for some $\bs{\rho}\in A$. Since the
one-site marginal $\nu_{\bs{\rho}}^1$ has some exponential moments for $\bs{\rho}\in A$, by Cram\'{e}r's theorem, the large
deviations functional of the family $\{\bs{\eta}(x)\}_{x\in\ZZ^d}$ is given by the Legendre transform $\Lambda_{\bs{\rho}}^*$ of
the logarithmic moment-generating functional $\Lambda_{\bs{\rho}}$. Note that~\eqref{LMGF} implies that modulo an affine function
depending on $\bs{\rho}$, the rate functional $\Lambda_{\bs{\rho}}^*$ coincides with the thermodynamic entropy $S$, that is
\begin{align}
  \label{RFviaTE}
  \Lambda_{\bs{\rho}}^*(\bs{\lambda})=S(\bs{\lambda})-\big\ls\bs{\lambda},\log\bs{\Phi}(\bs{\rho})\big\rs+\log Z\big(\bs{\Phi}(\bs{\rho})\big).
\end{align} 

\begin{lemma}
  \label{LDPlemma} 
  Let $\bs{\rho}\colon\T^d\to\bs{R}(\DD_{\bs{R}}^o)$ be a continuous profile and let $G\colon\T^d\x\RR_+\to\RR$ be a continuous
  function such that
  \begin{align}
    \label{GIneq}
    \sup_{u\in\T^d}|G(u,\bs{\lambda})|\leq C(1+|\bs{\lambda}|_1)\quad\text{for all }\bs{\lambda}\in\RR_+^2
  \end{align}
  for some constant $C>0$ such that $(2C,2C)\in\big(\bigcap_{u\in\T^d}\DD_{\Lambda_{\bs{\rho}(u)}}\big)^o$. Then 
  \begin{equation*}
    \limsup_{\ell\ra\infty}\limsup_{N\ra\infty}\fr{N^d}\log\int e^{\sum_{x\in\T_N^d}G(\frac{x}{N},\bs{\eta}^\ell(x))}d\nu_{\bs{\rho}(\cdot)}^N\leq
    \int_{\T^d}\sup_{\bs{\lambda}\in\RR_+^2}\Big\{G(u,\bs{\lambda})-\fr{2}\Lambda_{\bs{\rho}(u)}^*(\bs{\lambda})\Big\}du.
  \end{equation*}
\end{lemma}

We omit the proof of this Lemma as it is a simple adaptation of the corresponding result in the one-species case, \cite[Lemma
  6.1.10]{Kipnis1999a}. By the bound~\eqref{UnifBoundOnGOnFinTImeHorizon} the function
$G\colon[0,T_{\rm{max}})\x\T^d\x\RR_+^2\to\RR$ defined in~\eqref{G} satisfies
  \begin{equation*}
    \sup_{u\in\T^d}|G_t(u,\bs{\lambda})|\leq C_t(1+|\bs{\lambda}|_1)
  \end{equation*}
  for each fixed $t\in[0,T_{\rm{max}})$. Therefore, if we choose $\g_t>0$ small enough so that
  $2\g_t C_t(\bs{e}_1+\bs{e}_2)\in\big(\bigcap_{u\in\T^d}\Lambda_{\bs{\rho}_t(u)}\big)^o$, then for all $\g\in(0,\g_t)$ the
  function $\g G_t$ satisfies the assumptions of Lemma~\ref{LDPlemma}, and thus for $\g\in(0,\g_t)$ the term
  in~\eqref{AlmostGronwall} is bounded above by
\begin{align}
  \label{ToCompleteRelEntrMethod}
  \int_{\T^d}\sup_{\bs{\lambda}\in\RR_+^2}\Big\{\g G_t(u,\bs{\lambda})-\fr{2}\Lambda^*_{\bs{\rho}_t(u)}(\bs{\lambda})\Big\}du.
\end{align}
To complete the application of the relative entropy method, it remains to show that by reducing $\g_t>0$, $t\in[0,T_{\rm{max}})$,
  if necessary, this last term is non-positive.

We note that this would follow if we had a bound of the form
\begin{align}
  \label{TCREM}
  B_t:=\sup_{\substack{\bs{\rho}\in K_t \\\bs{\lambda}\in\RR_+^2}}
  \frac{|\bs{\Psi}(\bs{\rho},\bs{\lambda})|}{\Lambda_{\bs{\rho}}^*(\bs{\lambda})}<+\infty,
\end{align}
where $K_t\subs A:=\bs{R}(\DD_{\bs{R}}^o)\cap(0,\infty)^2$ is a compact set containing the image $\bs{\rho}_t(\T^d)$. Indeed,
since $\Lambda_{\bs{\rho}}^*(\bs{\lambda})=0$ iff $\bs{\lambda}=\bs{\rho}$, in which case $\bs{\Psi}(\bs{\rho},\bs{\lambda})=0$,
we would then have that
\begin{equation*}
  |\bs{\Psi}(\bs{\rho},\bs{\lambda})|\leq B_t\Lambda_{\bs{\rho}}^*(\bs{\lambda})
  \quad\text{ for all }(\bs{\rho},\bs{\lambda})\in K_t\x\RR_+^2,
\end{equation*} 
and so for $\g\in(0,\g_t)$ we would have 
\begin{equation*}
  \g|G_t(u,\bs{\lambda})|\leq\g \left\|\frac{\D[\bs{\Phi}(\bs{\rho}_t)]}{\bs{\Phi}(\bs{\rho}_t)}\right\|_{L^\infty(\T^d;\ell^2_\infty)}
  B_t\Lambda^*_{\bs{\rho}_t(u)}(\bs{\lambda})
\end{equation*}
for all $(u,\bs{\lambda})\in\T^d\x\RR_+^2$. Then by choosing $\g_t>0$ small enough so that in addition
$\g_tB_t\left\|{\D[\bs{\Phi}(\bs{\rho}_t)]}/{\bs{\Phi}(\bs{\rho}_t)}\right\|_{L^\infty(\T^d;\ell^2_\infty)}<\fr{2}$, it would
follow that~\eqref{ToCompleteRelEntrMethod} is non-positive, and the proof would be complete. The bound~\eqref{TCREM} is proved in
Lemma~\ref{LastBound}. Before we proceed with the proof of Lemma~\ref{LastBound}, we recall some facts on recession functions of
convex functions.

Given a lower semicontinuous proper convex function $\psi\colon\RR^d\to(-\infty,+\infty]$ with $0\in\DD_\psi$, its \emph{recession
  function} $\psi_\infty\colon\RR^d\to(-\infty,+\infty]$ is defined by
\begin{equation*}
  \psi_\infty(y):=\lim_{t\ra+\infty}\frac{\psi(ty)}{t}=\lim_{t\ra+\infty}\frac{d}{dt}\Big|_+\psi(ty),
\end{equation*}
where $\frac{d}{dt}|_+$ denotes differentiation from the right. The recession function $\psi_\infty$ is obviously positively
$1$-homogeneous, $\psi_\infty(\lambda y)=\lambda\psi_\infty(y)$ for all $y\in\RR^d$, $\lambda\geq 0$.

It is well known~\cite[Theorem 8.5]{Rockafellar1997a} that if $\psi$ is a proper lower semi-continuous convex function, then so is
its recession function. Using the equivalent definition of recession functions via the recession cone of their
epigraphs~\cite[Section 8]{Rockafellar1997a}, one can express the recession function by the formula
\begin{align}
  \label{FrP}
  \psi_\infty(y)=\inf\Big\{\liminf_{k\ra+\infty}\frac{\psi(t_ky_k)}{t_k}\Big|t_k\ra+\infty,\;y_k\ra y\Big\}
\end{align} 
(see~\cite[(12.7.1)]{Facchinei2003a}). Particularly useful in the proof of the following lemma is the characterisation of the
interior of the proper domain of a convex function $\psi$ via the recession function of its Legendre transform, as stated
in~\cite[(12.7.3)]{Facchinei2003a},
\begin{align}
  \label{IntDomViaLegRec}
  \DD_\psi^o=\bigcap_{y\neq 0}\big\{x\in\RR^d\bigm|\ls x,y\rs<(\psi^*)_\infty(y)\big\}.
\end{align}
Applying~\eqref{IntDomViaLegRec} to the \emph{thermodynamic pressure} $P:=\log\Z$, $\Z:=Z\circ\exp$, we get
\begin{equation}
  \label{PartDomCharRecess}
  \log\big(\DD_Z^o\cap(0,\infty)^2\big)=\DD_\Z^o
  =\big\{\bs{\mu}\in\RR^2\bigm|S_\infty(\bs{\lambda})>\ls\bs{\lambda},\bs{\mu}\rs,\;\forall\bs{\lambda}\neq0\big\}.
\end{equation}
In other words, $\DD_\Z^o$ is the intersection of all hyperplanes $\{\bs{\mu}\in\RR^2|\ls\bs{\mu},\bs{\y}\rs<S_\infty(\bs{\y})\}$
for $\bs{\y}\in S^1\cap(0,\infty)^2$. This implies that the function
$S^1\cap(0,\infty)^2\ni\bs{\y}\mapsto S_\infty(\bs{\y})\bs{\y}\in\RR^2$ is a parametrisation of the boundary $\pd\DD_\Z$. This may
be compared with~\cite[(2.14)]{Grosskinsky2008a}. Consequently, the part of the boundary $\pd\DD_Z$ on the strictly positive
quadrant is given by the parametrisation $e^{S_\infty(\bs{\y})\bs{\y}}$, $\bs{\y}\in S^1\cap(0,\infty)^2$.  Along the two axes
$\phih_1=0$ and $\phih_2=0$, there is only one-species of particles and the critical fugacities in these directions are fugacities
of one species ZRPs.

\begin{lemma}
  \label{LastBound} 
  For any compact $K\subs A:=\bs{R}(\DD_{\bs{R}}^o)\cap(0,\infty)^2$,
  \begin{equation}
    \label{eq:hamster}
    \sup_{(\bs{\rho},\bs{\lambda})\in K\x\RR_+^2}\frac{|\bs{\Psi}(\bs{\rho},\bs{\lambda})|}{\Lambda_{\bs{\rho}}^*(\bs{\lambda})}<+\infty.
  \end{equation}
\end{lemma}

\proof For all $(\bs{\rho},\bs{\lambda})\in K\x(0,\infty)^2$ we have that
$\Lambda^*(\bs{\rho},\bs{\lambda}):=\Lambda^*_{\bs{\rho}}(\bs{\lambda})\geq 0$ and the functions $|\bs{\Psi}| \colon
K\x(0,\infty)^2\to\RR_+$ and $\Lambda^* \colon K\x(0,\infty)^2\to\RR_+$ are continuous. Therefore the fraction in the supremum can
tend to infinity if the nominator goes to infinity or the denominator goes to zero. Since $\bs{\Psi} \colon
K\x(0,\infty)^2\to\RR^2$ is continuous and $K$ is compact the nominator can tend to infinity only as $|\bs{\lambda}|_1\ra+\infty$.
In this case $\bs{\Lambda}_{\bs{\rho}}^*$ also tends to $+\infty$ as a rate functional with compact level sets. Since
$\Lambda_{\bs{\rho}}^*$ is the rate functional of the i.i.d.~occupation variables $\bs{\eta}(x)$, $x\in\ZZ^d$, with common law
$\nu_{\bs{\rho}}^1$ we have that $\Lambda^*(\bs{\rho},\bs{\lambda})=0$ iff $\bs{\rho}=\bs{\lambda}$ for the denominator. But
obviously for $\bs{\rho}=\bs{\lambda}$ we have $\bs{\Psi}(\bs{\rho},\bs{\lambda})=0$, so the nominator vanishes as well. So in
order to prove the lemma we have to show that the nominator and the denominator are of the same order as
$|\bs{\rho}-\bs{\lambda}|\ra 0$ and $|\bs{\rho}-\bs{\lambda}|\ra+\infty$.

Motivated by the previous sketch, we choose $\ee>0$ such that $K_\ee:=\bbar{K^{(\ee)}}\subs A$, where
$K^{(\ee)}:=\bigcup_{x\in K}D(x,\ee)$, and for any $M>0$ we separate the region $K\x(0,\infty)^2$ as
$K\x(0,\infty)^2=\E_0^\ee\cup\E_\ee^M\cup\E_M^\infty$, where
\begin{align*}
  \E_0^\ee &:=\{(\bs{\rho},\bs{\lambda})\in K\x(0,\infty)^2\bigm||\bs{\rho}-\bs{\lambda}|\leq\ee\},\\
  \E_\ee^M &:=\{(\bs{\rho},\bs{\lambda})\in K\x(0,\infty)^2\bigm|\ee\leq|\bs{\rho}-\bs{\lambda}|\leq M\},\\
  \E_M^\infty &:=\{(\bs{\rho},\bs{\lambda})\in K\x(0,\infty)^2\bigm||\bs{\rho}-\bs{\lambda}|\geq
                M\}.
\end{align*}
We prove the claim on each region individually. Obviously the set $\E_\ee^M$ is compact and so since the functions $\bs{\Psi}$ and
$\Lambda^*$ are jointly continuous, the claim holds on the region $\E_\ee^M$.

We turn to the region $\E_0^\ee$. By its definition, for any $(\bs{\rho},\bs{\lambda})\in\E_0^\ee$ we have that
$\bs{\lambda}\in D(\bs{\rho},\ee)\subs K_\ee\subs A$. So, since $D(\bs{\rho},\ee)$ is convex, for all
$(\bs{\rho},\bs{\lambda})\in\E_0^\ee$ the image of the constant speed line segment
$\bs{\g}_{\bs{\rho},\bs{\lambda}}\colon[0,1]\to\RR^2$ from $\bs{\rho}$ to $\bs{\lambda}$ is contained in $K_\ee$, i.e.,
\begin{align}
  \label{LocConv}
  \bs{\g}_{\bs{\rho},\bs{\lambda}}([0,1])\subs K_\ee\quad\text{ for all }(\bs{\rho},\bs{\lambda})\in\E_0^\ee.
\end{align} 
By the first order Taylor expansion of $\Phi_i$, $i=1,2$ around the point $\bs{\rho}\in K$,
\begin{equation*}
  \Psi_i(\bs{\rho},\bs{\lambda})=\int_0^1(1-t)\big\ls\bs{\lambda}
  -\bs{\rho},D^2\Phi_i(\bs{\g}_{\bs{\rho},\bs{\lambda}}(t))(\bs{\lambda}-\bs{\rho})\big\rs dt
  \quad\text{for all }(\bs{\rho},\bs{\lambda})\in\E_0^\ee.
\end{equation*} 
Since $\bs{\Phi}$ is smooth on the set $A$, the matrix $D^2\Phi_i(\bs{\rho})$ is symmetric for all $\bs{\rho}\in A$. Denoting by
$\lambda_{\pm}^i(\bs{\rho})$ the real eigenvalues of $D^2\Phi_i(\bs{\rho})$ we have
\begin{equation*}
  \lambda_-^i(\bs{\g}_{\bs{\rho},\bs{\lambda}}(t))|\bs{\lambda}-\bs{\rho}|^2\leq\ls\bs{\lambda}-\bs{\rho},D^2
  \Phi_i(\bs{\g}_{\bs{\rho},\bs{\lambda}}(t))(\bs{\lambda}-\bs{\rho})\big\rs\leq\lambda_+^i(\bs{\g}_{\bs{\rho},\bs{\lambda}}(t))
  |\bs{\lambda}-\bs{\rho}|^2.
\end{equation*} 
Furthermore, by the continuity of the eigenvalues $\lambda^i_\pm$ as functions of $\bs{\rho}\in A$,
\begin{equation*}
  A^i:=\sup_{\bs{\rho}\in K_\ee}|\lambda^i_-|\mx|\lambda^i_+|(\bs{\rho})<+\infty.
\end{equation*} 
So by~\eqref{LocConv}, we have that $|\lambda^i_-|\mx|\lambda^i_+|(\bs{\g}_{\bs{\rho},\bs{\lambda}}(t))\leq A^i$ for all
$(t,\bs{\rho},\bs{\lambda})\in[0,1]\x\mathcal{E}_0^\ee$ and thus
\begin{equation*}
  |\Psi_i(\bs{\rho},\bs{\lambda})|\leq\frac{A^i}{2}|\bs{\lambda}-\bs{\rho}|^2,\quad i=1,2.
\end{equation*} 
For the denominator in~\eqref{eq:hamster}, we note that the rate functional $\Lambda_{\bs{\rho}}^*$ is $C^1$ on $(0,\infty)^2$ and
$C^2$ on $A$ with
\begin{equation*}
  \nabla\Lambda_{\bs{\rho}}^*(\bs{\lambda})
  =\log\frac{\bar{\bs{\Phi}}(\bs{\lambda})}{\bs{\Phi}(\bs{\rho})},\quad\bs{\lambda}\in(0,\infty)^2,
\end{equation*}
\begin{equation*}
  D^2\Lambda_{\bs{\rho}}^*(\bs{\lambda})= D\log\bs{\Phi}(\bs{\lambda})=D^2S(\bs{\lambda}),\quad\bs{\lambda}\in A,
\end{equation*}
where $S$ is the thermodynamic entropy. Since $\Lambda_{\bs{\rho}}^*$ and its derivative vanish at $\bs{\rho}$, by Taylor
expansion of $\Lambda_{\bs{\rho}}^*$ around $\bs{\rho}\in K$
\begin{equation*}
  \Lambda_{\bs{\rho}}^*(\bs{\lambda})=\int_0^1(1-t)\big\ls\bs{\lambda}
  -\bs{\rho},D^2S(\bs{\g}_{\bs{\rho},\bs{\lambda}}(t))(\bs{\lambda}
  -\bs{\rho})\big\rs dt,\quad\lambda\in A.
\end{equation*} 
Denoting by $\lambda_-(\bs{\rho})>0$ the minimal eigenvalue of the strictly positive definite matrix $D^2S(\bs{\rho})$, we have by
continuity that
\begin{equation*}
  B:=\inf_{\bs{\rho}\in K_\ee}\lambda_-(\bs{\rho})>0.
\end{equation*} 
Then $\Lambda_{\bs{\rho}}^*(\bs{\lambda})\geq\frac{B}{2}|\bs{\lambda}-\bs{\rho}|^2$ for all $(\bs{\rho},\bs{\lambda})\in\E_0^\ee$,
which shows that
\begin{equation*}
  \sup_{(\bs{\rho},\bs{\lambda})\in\E_0^\ee}\frac{|\Psi_i(\bs{\rho},\bs{\lambda})|}{\Lambda_{\bs{\rho}}(\bs{\lambda})}\leq \frac{A^i}{B}<+\infty,\quad i=1,2
\end{equation*} 
and yields the bound~\eqref{eq:hamster} in the region $\mathcal{E}_0^\ee$.

It remains to show that the supremum is finite in the region $\mathcal{E}_M^\infty$ for some $M>0$. On one hand, it follows
from~\eqref{PhiQuasiLip} and the compactness of $K$ that $\bs{\Psi}$ satisfies a bound of the form
\begin{equation*}
  |\bs{\Psi}(\bs{\rho},\bs{\lambda})|_1\leq C_0+C_1|\bs{\lambda}|_1\quad\forall\;(\bs{\rho},\bs{\lambda})\in K\x\RR_+^2
\end{equation*}
for some constants $C_0,C_1\geq 0$. So, to complete the proof, it suffices to show that $\Lambda^*$ has at least linear growth in
$\E_M^\infty$ as $|\bs{\lambda}|\ra+\infty$, i.e.,
\begin{equation*}
  \lim_{M\ra+\infty}\inf_{(\bs{\rho},\bs{\lambda})\in\E_M^\infty}\frac{\Lambda^*_{\bs{\rho}}(\bs{\lambda})}{|\bs{\lambda}|}>0,
\end{equation*} 
where of course the limit as $M\ra+\infty$ exists as an increasing limit. We begin by noting that
\begin{equation*}
  \lim_{M\ra+\infty}\inf_{(\bs{\rho},\bs{\lambda})\in\E_M^\infty}\frac{\Lambda^*_{\bs{\rho}}(\bs{\lambda})}{|\bs{\lambda}|}
  \geq \liminf_{|\lambda|\ra+\infty}\inf_{\bs{\rho}\in K}\frac{\Lambda_{\bs{\rho}}^*(\bs{\lambda})}{|\bs{\lambda}|}=:a.
\end{equation*}
We choose a sequence $\{\bs{\lambda}_n\}\subs\RR_+^2$ achieving the limit inferior,
\begin{equation*}
  |\bs{\lambda}_n|\ra+\infty\quad\text{and}\quad\lim_{n\ra+\infty}\inf_{\bs{\rho}\in K}
  \frac{\Lambda_{\bs{\rho}}^*(\bs{\lambda}_n)}{|\bs{\lambda}_n|}=a.
\end{equation*} 
Since $\{\frac{\bs{\lambda}_n}{|\bs{\lambda}_n|}\}$ is contained in the compact space $S^1_+:=S^1\cap\RR_+^2$, by passing to a
subsequence if necessary, we can assume that $\{\frac{\bs{\lambda}_n}{|\bs{\lambda}_n|}\}$ converges to some direction
$\bs{\y}\in S^1_+$. Then obviously
\begin{align}
  \label{Greater}
  \liminf_{n\ra+\infty}\frac{S(\bs{\lambda}_n)}{|\bs{\lambda}_n|}\geq  
  \liminf_{\substack{|\bs{\lambda}|\ra+\infty\\{\bs{\lambda}}/{|\bs{\lambda}|}\ra\y}}\frac{S(\bs{\lambda})}{|\bs{\lambda}|}
  =S_\infty(\bs{\y}),
\end{align}
where the equality in the right-hand side holds by~\eqref{FrP}.  Since $\bs{\Phi}(K)\subs\DD_Z^o\cap(0,\infty)^2$, we have
by~\eqref{PartDomCharRecess} that $S_\infty(\bs{\y})-\ls\bs{\y},\log\bs{\Phi}(\bs{\rho})\rs>0$ for all $\bs{\rho}\in K$. Thus,
since $\bs{\Phi}$ is continuous and $K$ is compact,
\begin{equation*}
  \theta:=\inf_{\bs{\rho}\in K}\{S_\infty(\bs{\y})-\ls\bs{\y},\log\bs{\Phi}(\bs{\rho})\}>0.
\end{equation*} 
Then by~\eqref{Greater} there exists $n_1\in\NN$ such that
\begin{equation*}
  n\geq n_1\quad\Longrightarrow\quad \frac{S(\bs{\lambda}_n)}{|\bs{\lambda}_n|}\geq S_\infty(\bs{\y})-\frac{\theta}{3}.
\end{equation*}
By~\eqref{RFviaTE} and taking into account the fact that $Z\geq 1$, we have that for all $n\geq n_1$ and all $\bs{\rho}\in K$,
\begin{align*}
  \frac{\Lambda_{\bs{\rho}}^*(\bs{\lambda}_n)}{|\bs{\lambda}_n|}
  &\geq S_\infty(\bs{\y})-\Big\ls\frac{\bs{\lambda}_n}{|\bs{\lambda}_n|},\log\bs{\Phi}(\bs{\rho})\Big\rs
    +\fr{|\bs{\lambda}_n|}\log Z\big(\bs{\Phi}(\bs{\rho})\big)-\frac{\theta}{3}\\
  &\geq S_\infty(\bs{\y})-\Big\ls\frac{\bs{\lambda}_n}{|\bs{\lambda}_n|},\log\bs{\Phi}(\bs{\rho})\Big\rs-\frac{\theta}{3}.
\end{align*}
But by the compactness of $K$, we have that
$\|\log\bs{\Phi}\|_{L^\infty(K)}:=\sup_{\bs{\rho}\in K}|\log\bs{\Phi}(\bs{\rho})|_2<+\infty$ and therefore the sequence
$\{\ls\frac{\bs{\lambda}_n}{|\bs{\lambda}_n|},\log\bs{\Phi}(\bs{\rho})\rs\}$ converges to $\ls\bs{\y},\log\bs{\Phi}(\bs{\rho})\rs$
uniformly over all $\bs{\rho}\in K$,
\begin{equation*}
  \sup_{\bs{\rho}\in K}\Big|\Big\ls\frac{\bs{\lambda}_n}{|\bs{\lambda}_n|},
  \log\bs{\Phi}(\bs{\rho})\Big\rs-\big\ls\bs{\y},\log\bs{\Phi}(\bs{\rho})\big\rs\Big|
  \leq\|\log\bs{\Phi}\|_{L^\infty(K)}\Big|\frac{\bs{\lambda}_n}{|\bs{\lambda}_n|}-\bs{\y}\Big|_2\to
  0.
\end{equation*} 
Therefore we can choose $n_2\in\NN$ such that
\begin{equation*}
  n\geq n_2\quad\Longrightarrow\quad\sup_{\bs{\rho}\in K}\Big|\Big\ls\frac{\bs{\lambda}_n}{|\bs{\lambda}_n|},
  \log\bs{\Phi}(\bs{\rho})\Big\rs
  -\big\ls\bs{\y},\log\bs{\Phi}(\bs{\rho})\big\rs\Big|<\frac{\theta}{3},
\end{equation*}
and then for all $n\geq n_1\mx n_2$ and all $\bs{\rho}\in K$
\begin{equation*}
  \frac{\Lambda_{\bs{\rho}}^*(\bs{\lambda}_n)}{|\bs{\lambda}_n|}\geq
 S_\infty(\bs{\y})-\Big\ls\bs{\y},\log\bs{\Phi}(\bs{\rho})\Big\rs-\frac{2\theta}{3}
 \geq\theta-\frac{2\theta}{3}=\frac{\theta}{3}>0.
\end{equation*} 
This proves that 
\begin{equation*}
  \liminf_{|\bs{\lambda}|\ra+\infty}\inf_{\bs{\rho}\in K}\frac{\Lambda_{\bs{\rho}}^*(\bs{\lambda})}{|\bs{\lambda}|}
  =\lim_{n\ra+\infty}\inf_{\bs{\rho}\in K}\frac{\Lambda_{\bs{\rho}}^*(\bs{\lambda}_n)}{|\bs{\lambda}_n|}>0,
\end{equation*} which completes the proof of Lemma~\ref{LastBound}. \qed
     		
Since Lemma~\ref{LastBound} establishes the missing bound~\eqref{TCREM}, the proof of Theorem~\ref{HL} is complete.

\subsection{Proof of Theorem~\protect{\ref{PDE}}}
\label{sec:Proof-Theor}

By~\cite{Amann1986a}, it is known that quasilinear parabolic systems have unique maximal classical solutions when starting from
initial profiles of class $C^{2+\theta}$, $\theta\in[0,1)$. To show that classical solutions of the blind-species
  system~\eqref{SumSystemPDE} are global in time, we prove first a maximum principle asserting that the region
\begin{equation*}
  A:=\bs{R}(\DD_{\bs{R}}^o)\cap(0,\infty)^2=\big\{\bs{\rho}\in(0,\infty)^2\bigm|\rho_1+\rho_2<\hat{\rho}_c\big\}
\end{equation*}
is invariant under the evolution of the species-blind system. Here $\hat{\rho}_c\in(0,+\infty]$ is the critical density of the
$1$-species ZRP associated with the species-blind ZRP. The proof of this version of the maximum principle for systems of the
form~\eqref{SumSystemPDE} relies on the maximum principle for quasilinear PDEs in divergence form found
in~\cite{Arena1972a}. Since, as we will see, the solution $\bs{\rho}$ cannot lose regularity, we will obtain the existence
of global in time classical solutions.

\begin{lemma}[A weak maximum principle for the species-blind system] 
  Let $\bs{\rho} = (\rho_1,\rho_2)\in C^{1,2}([0,T)\x\T^d;\RR^2)$, $T > 0$, be a classical solution to the
  problem~\eqref{SumSystemPDE} starting from an initial condition $\bs{\rho}_0\in C(\T^d;\RR^2_+)$ satisfying
  \begin{equation}
    \label{InitCond}
    \bs{\rho}_0(\T^d)\subs A = \big\{\bs{\rho}\in(0,\infty)^2\bigm|\rho_1 + \rho_2 < \hat{\rho}_c\big\},
  \end{equation}
  where $\hat{\rho}_c$ is the critical density corresponding to the $1$-species density function $\hat{R}$. Then
  \begin{equation}
    \label{MP} 0 < \inf_{(t,u)\in[0;T)\x\T^d}\rho_1(t,u)\mn\rho_2(t,u)\leq\sup_{(t,u)\in[0,T)\x\T^d}
    \big(\rho_1(t,u)+\rho_2(t,u)\big) <\hat{\rho}_c.
  \end{equation}
  In particular, for some $\delta>0$, $\bs{\rho}_t(\T^d)\subs \{\bs{r}\in A\bigm|d(\bs{r},\pd A) > \delta\}$ for all $t\in[0,T)$.
\end{lemma}

\proof By the continuity of $\bs{\rho}_0$ and the compactness of $\T^d$, there exists by assumption~\eqref{InitCond} an $\ee>0$
such that
\begin{equation}
  \label{eeInitCond}
  \bs{\rho}_0(\T^d)\subs\{(\rho_1,\rho_2)\in\RR^2 \bigm| \rho_1\mn\rho_2 > \ee,\rho_1+\rho_2< \hat{\rho}_c-\ee\},
\end{equation}
where we replace $\hat{\rho}_c-\ee$ by $\fr{\ee}$ when $\hat{\rho}_c=+\infty$. Since $\bs{\rho}$ solves~\eqref{SumSystemPDE}, by
summing the two equations we see that the function $\rho_1+\rho_2$ solves the equation $\pd_t\rho = \Delta\hat{\Phi}(\rho)$. But
since $\hat{\Phi}$ is the mean jump rate of a single species ZRP,
\begin{align}
  \label{UnifPar} 0 < c <\hat{\Phi}'(\rho) < C < +\infty\quad\text{ for all } \rho\in[0,\hat{\rho}_c-\ee/2]
\end{align}
for some constants $c,C \geq 0$ and therefore the equation $\pd_t\rho = \Delta\hat{\Phi}(\rho)$ is uniformly parabolic, when
considered for sub-critical initial conditions $\rho_0\in C(\T^d,(0,\hat{\rho}_c))$. Therefore it follows by~\eqref{eeInitCond}
and the maximum principle for scalar uniformly parabolic quasilinear equations that 
\begin{equation}
  \label{MPforSum}
  2\ee<\inf_{(t,u)\in[0;T)\x\T^d}(\rho_1+\rho_2)(t,u)\leq\sup_{(t,u)\in[0;T)\x\T^d}(\rho_1+\rho_2)(t,u)<\hat{\rho}_c-\ee.
\end{equation}
We consider now the family of the open domains
\begin{equation*}
  B_\delta:=\{(\rho_1,\rho_2)\in\RR^2|\ee<\rho_1+\rho_2<\hat{\rho}_c-\ee,\;\rho_1\mn\rho_2>-\delta\}
\end{equation*}
for $\delta\in[0,+\infty]$ and set
\begin{equation*}
  D_\delta:=\{(t,u,r)\in[0,T)\x\T^d\x\RR|(r,\rho_2(t,u))\in B_\delta\}.
\end{equation*}
Let $\Psi\colon D_\infty\to\RR$ denote the function given by the formula
\begin{equation*}
  \Psi(t,u,r)=r\frac{\hat{\Phi}(r+\rho_2(t,u))}{r+\rho_2(t,u)}.
\end{equation*}
The sets $D_\delta$ are obviously open and the function $\Psi$ is well defined on $D_\infty$. Since the sum $\rho_1+\rho_2$
satisfies~\eqref{MPforSum}, we have that
\begin{equation*}
  (t,u,\rho_1(t,u))\in D_\infty\quad\text{for all }(t,u)\in[0,T)\x\T^d,
\end{equation*}
and since $(\rho_1,\rho_2)$ is a solution of~\eqref{SumSystemPDE}, we have that $\rho_1$ solves
\begin{equation*}
  \pd_t\rho_1(t,u)=\Delta\Psi\big(t,u,\rho_1(t,u)\big).
\end{equation*}
In divergence form, the problem above is written as
\begin{align}
  \label{DivForm}
  \pd_t\rho_1(t,u)=\dv A_\Psi\big(t,u,\rho_1(t,u),\nabla \rho_1(t,u)\big)
\end{align}
where $A_\Psi\colon D_\infty\to\RR^d$ is the function given by the formula
\begin{equation*}
  A_\Psi(t,u,r,\y)=\nabla_u\Psi(t,u,r)+\pd_r\Psi(t,u,r)\y.
\end{equation*}
Since $\rho_2$ is $C^{1,2}$, it follows that the function $A_\Psi$ is $C^1$ and $\pd_\y A_\Psi(t,u,r,\y)=\pd_r\Psi(t,u,r)I$ where
$I\in\RR^{d\x d}$ denotes the identity matrix. By a simple calculation, $\pd_r\Psi(t,u,r)=H(r,\rho_2(t,u))$, where $H\colon
B_\infty\to\RR$ is given by
\begin{equation*}
  H(\rho_1,\rho_2)=\frac{\rho_2}{\rho_1+\rho_2}\frac{\hat{\Phi}(\rho_1+\rho_2)}{\rho_1+\rho_2}
  +\frac{\rho_1}{\rho_1+\rho_2}\hat{\Phi}'(\rho_1+\rho_2).
\end{equation*}
We have that
\begin{align}
  \label{HB}
  \inf_{B_\delta}H\leq\inf_{D_\delta}\pd_r\Psi\leq\sup_{D_\delta}\pd_r\Psi\leq\sup_{B_\delta}H
\end{align} 
for all $\delta\in[0,+\infty]$ and it is obvious that 
\begin{equation*}
  c\leq\inf_{B_0}H\leq\sup_{B_0}H\leq C,
\end{equation*} 
where $c,C\geq 0$ are the constants in~\eqref{UnifPar}. By continuity of $H$, we obtain the existence of $\delta_0 > 0$ such that
\begin{align}
  \label{HBD}
  \frac{c}{2}<\inf_{B_{\delta_0}}H\leq\sup_{B_{\delta_0}}H<2C,
\end{align}
which shows that the diagonal matrix $\pd_\y A_\Psi$ is positive definite on the set $D_{\delta_0}\x\RR^d$.  We set now
\begin{equation*}
  T^i:=\sup\Big\{t\in[0,T]\Big|\inf_{(s,u)\in[0,t)\x\T^d}\rho_i(s,u)>0\Big\},\quad i=1,2.
\end{equation*}
By the assumptions on the initial condition $\bs{\rho}_0$, the set over which we take the supremum is non-empty. By the continuity
of the solution $\bs{\rho}$, we have $T^i > 0$ for $i = 1, 2$ and if $T^i < T$ then there exists $u_0^i\in\T^d$ such that
$\rho_i(T^i,u_0^i) = 0$. In order to prove the claim of the lemma, it suffices to show that $T^1 = T^2 = T$.

So we suppose that this is not true to obtain a contradiction. Without loss of generality it suffices to consider the cases
$T^1 < T^2 < T$ and $T^0 := T^1 = T^2 < T$.

(a) $T^1<T^2<T$: Since $\rho_1(t,u)\geq 0$ for all $(t,u)\in[0,T^1]\x\T^d$ and $\rho_1$ is continuous in $[0,T)\x\T^d$, there
exists $t_0 > 0$ such that
\begin{equation*}
  \inf_{(t,u)\in[0,T^1+t_0]\x\T^d}\rho_1(t,u) > -\delta_0.
\end{equation*}
But then $(t,u,\rho_1(t; u))\in D_{\delta_0}$ for all $(t,u) \in [0,T^1 + t_0]\x\T^d$ and so, since $\rho_1$ and $0$ are solutions
of problem~\eqref{DivForm} in $[0, T^1 + t_0] \x \T^d$, which is uniformly parabolic in this region by~\eqref{HB} and~\eqref{HBD},
and since $\rho_1(T^1,u^1_0) = 0$, we get from~\cite[Theorem 1]{Arena1972a} that $\rho_1 \equiv 0$ in $[0,T^1) \x \T^d$, which
contradicts the definition of $T^1$.

(b) $T^0 := T^1 = T^2 < T$: Again, since $\rho_1(t,u)\mn\rho_2(t,u)>0$ for all $(t,u)\in[0,T^0]\x\T^d$, there exists $t_0 > 0$
such that
\begin{equation*}
  \inf_{(t,u)\in[0,T^1+t_0]\x\T^d}[\rho_1(t,u)\mn\rho_2(t,u)]\geq-\delta_0.
\end{equation*}
But then again the problem~\eqref{DivForm} is uniformly parabolic in $[0,T^0 +t_0]\x\T^d$ and $\rho_1$ and $0$ are solutions with
$\rho_1 \geq 0$ in $[0,T^0]$, which again by~\cite[Theorem 1]{Arena1972a} yields $\rho_1\equiv 0$ in $[0,T^0)\x\T^d$ and
contradicts the definition of $T^0$. \qed
     	   
Using this maximum principle and the global existence of scalar uniformly parabolic equations, we obtain the global existence of
solutions to the species-blind system as follows. To derive a contradiction, we assume that $\bs{\rho}\in
C^{1,2}([0,T_{\rm{max}})\x\T^d;\RR^2)$, $T_{\rm{max}}<+\infty$, is the maximal classical solution of the species-blind system
  starting from $\bs{\rho}_0$. Here maximality of the solution means that $\bs{\rho}$ can not be extended to a $C^{1,2}$-solution
  on $[0,T]\x\T^d$ for $T>T_{\rm{max}}$. Since $\hat{\rho}_0 := \rho_{01}+\rho_{02}\in C^{1+\theta;2+\theta}(\T^d,(0,\rho_c))$,
  there exists a unique solution $\hat{\rho}\in C^{1+\theta,2+\theta}(\RR_+\x\T^d;(0,\rho_c))$ of the scalar quasilinear parabolic
  equation $\pd_t\rho = \Delta\hat{\Phi}(\rho)$ with initial data $\hat{\rho}_0$.  Then, since
  $\hat{\rho}(\RR_+\x\T^d)\subs(\ee,\hat{\rho}_c-\ee)$ for some $\ee > 0$ and the function $\phi(x) := \frac{\hat{\Phi}(x)}{x}$ is
  $C^\infty$ in $[\ee, \hat{\rho}_c-\ee]$, the function $a\colon\RR_+\x\T^d\to\RR_+$ defined by
  $a(t,u):=\frac{\hat{\Phi}(\hat{\rho}(t,u))}{\hat{\rho}(t,u)}$ belongs to $C^{1+\theta,2+\theta}(\RR_+\x\T^d)$. Since
  $\hat{\Phi}'$ satisfies~\eqref{UnifPar},
\begin{align}
  \label{aBounds}
  0<c<a(t,u)\leq C<+\infty\quad\text{for all }(t,u)\in\RR_+\x\T^d
\end{align}
for some constants $c,C\geq 0$. Since the function $\rho_1 + \rho_2$ is also a solution of the scalar equation
$\pd_t\rho = \Delta\hat{\Phi}(\rho)$ with the same initial data $\rho_0$, we have by the uniqueness of solutions that
\begin{align}
  \label{aThroughSolution}
  a\equiv\frac{\hat{\Phi}(\rho_1+\rho_2)}{\rho_1+\rho_2}\quad\text{in }[0,T_{\rm{max}})\x\T^d.
\end{align}

We consider the system
\begin{align}
  \label{Decoupled}
  \begin{cases}
    \pd_t\rho_1=\Delta\big(a(t,u)\rho_1(t,u)\big)\\
    \pd_t\rho_2=\Delta\big(a(t,u)\rho_2(t,u)\big)
  \end{cases},\quad
  \bs{\rho}(0,\cdot)=(\rho_{01},\rho_{02})\quad\text{in }\T^d,
\end{align}
which is obviously decoupled and can be solved by solving the scalar linear second order parabolic equation
\begin{align}
  \label{DecoupledScalar}
  \pd_t\rho=\Delta\big(a(t,u)\rho(t,u)\big)
\end{align}
twice with initial conditions $\rho_{01}$ and $\rho_{02}$. This scalar equation is given in general form by
\begin{equation*}
  \pd_t\rho=\sum_{i,j=1}^da^{ij}\pd_{ij}^2\rho+\sum_{i=1}^db^i\pd_i\rho+c\rho,
\end{equation*} 
where $a^{ij}=a\delta_{ij}$, $b^i=\pd_ia$ and $c=\Delta a$. We note that since $a$ satisfies~\eqref{aBounds} and $a^{ij} =
a\delta_{ij}$, the matrix $(a^{ij})$ is uniformly elliptic. Also, since $a\in C^{1+\theta,2+\theta}(\RR_+\x\T^d)$, the
coefficients $a^{ij},b^i,c$ are $\theta$-H\"{o}lder continuous and so by the interpretation of~\cite[Theorem 5.14]{Lieberman1996a}
in the flat torus with periodic boundary conditions, we find that for any $\rho_0 \in C^{2+\theta}(\T^d)$ there exists a unique
solution $\rho\in C^{1+\theta,2+\theta}_{\rm{loc}}(\RR_+\x\T^d)$ to the scalar problem~\eqref{DecoupledScalar} with initial
condition $\rho_0$, and thus there exists a unique solution $\wt{\bs{\rho}}\in
C^{1+\theta,2+\theta}_{\rm{loc}}(\RR_+\x\T^d;\RR^2)$ of system~\eqref{Decoupled} starting from $\bs{\rho}_0 =
(\rho_{01},\rho_{02})$. Since by~\eqref{aThroughSolution}, we have that the solution $\bs{\rho}\in
C^{1,2}([0,T_{\rm{max}})\x\T^d;\RR^2)$ of the system~\eqref{SumSystemPDE} also solves the system~\eqref{Decoupled}, it follows by
  the uniqueness of solutions that $\wt{\bs{\rho}}=\bs{\rho}$ in $[0,T_{\rm{max}})\x\T^d$. This, taking also into account the
    maximum principle, shows that
\begin{equation*}
  \bs{\rho}\in C^{1+\theta,2+\theta}([0,T_{\rm{max}})\x\T^d;A).
\end{equation*}

Now, we obviously have that $\wt{\bs{\rho}}_{T_{\max}}\in C^{2+\theta}(\T^d)$, and since $\wt{\bs{\rho}}$
solves~\eqref{SumSystemPDE} in $[0,T_{\rm{max}})\x\T^d$, we have by the maximum principle that
\begin{equation*}
  \wt{\bs{\rho}}([0,T_{\rm{max}})\x\T^d)\subs\{\bs{r}\in A|d(\bs{r},\pd A)>\delta\}
\end{equation*} 
for some $\delta>0$. Consequently, by continuity, we also have that $\wt{\bs{\rho}}_{T_{\rm{max}}}(\T^d)\subs A$. We consider then
a solution $\bs{r}\colon[0,\ee)\x\T^d\to A$, $\ee>0$, of the problem~\eqref{SumSystemPDE} starting from
$\bs{r}_0=\wt{\bs{\rho}}_{T_{\rm{max}}}$ and extend $\bs{\rho}$ on $[0,T_{\rm{max}}+\ee)\x\T^d$ by defining
$\bs{\rho}(t,\cdot):=\bs{r}(t-T_{\rm{max}},\cdot)$ for $t\in[T_{\rm{max}},T_{\rm{max}}+\ee)$. This function is obviously of class
$C^{1+\theta,2+\theta}$ and solves~\eqref{SumSystemPDE}, which contradicts the maximality of $T_{\rm{max}}$.  \qed

\subsection{Proof of Corollary~\protect{\ref{SpBlHL}}}
\label{sec:Proof-Coroll}

By the global existence in time of solutions to the species-blind parabolic system, it suffices to check that Theorem~\ref{HL}
applies. Since the the one-species partition function $\hat{Z}$ is continuous on $\DD_{\hat{Z}}$, it follows by the formula
$Z(\bs{\phih})=\hat{Z}(\phih_1+\phih_2)$ that the partition function is continuous. It remains to check that in the case where the
associated $1$-species ZRP has finite critical density, $\bs{g}$ has regular tails, i.e., that for every $\bs{\y}\in S^1_{1,+}$
\begin{align}
  \label{DirectionalCritFug}
  \mu_{c;1}(\bs{\y}):=\log\phih_{c;1}(\bs{\y})
  :=\liminf_{\substack{|\bs{k}|_1\ra+\infty\\\bs{k}/|\bs{k}|_1\ra\bs{\y}}}\fr{|\bs{k}|_1}\log\bs{g}!(\bs{k}),\quad\bs{\y}\in S^1_{1,+}
\end{align}
exists as a limit and is a continuous function of the direction $\bs{\y}\in S^1_{1,+}$. By the formula of $\bs{g}!$ we have that
\begin{align}
  \label{BlindTail}
  \fr{|\bs{k}|_1}\log\bs{g}!(\bs{k})=\fr{|\bs{k}|_1}\log\frac{k_1!k_2!}
  {|\bs{k}|_1!}+\fr{|\bs{k}|_1}\log \hat{g}!(|\bs{k}|_1).
\end{align}
The second term in the right hand side of~\eqref{BlindTail} converges as $|\bs{k}|_1\ra+\infty$ to the critical chemical potential
$\hat{\mu}_c=\log\hat{\phih}_c$ of the $1$-species jump rate $\hat{g}$. Since by Stirling's approximation
$\lim_{k\ra+\infty}\frac{k!}{\sqrt{2\pi k}(k/e)^k}=1$, we can replace the liminf of the first term in the right hand side
of~\eqref{BlindTail} by
\begin{align}
  \label{InteractChemPot}
  \liminf_{\substack{|\bs{k}|_1\ra+\infty\\\bs{k}/|\bs{k}|_1\ra\bs{\y}}}\fr{|\bs{k}|_1}\log\sqrt{2\pi}
  \frac{\sqrt{k_1}k_1^{k_1}\sqrt{k_2}k_2^{k_2}}{\sqrt{|\bs{k}|_1}|\bs{k}|_1^{|\bs{k}|_1}}.
\end{align} 
This limit inferior exists as a limit and defines a continuous function of $\bs{\y}$. Indeed, for all $\bs{k}\in\NN^2$
we have that
\begin{equation*}
  \fr{|\bs{k}|_1}\log\frac{\sqrt{k_1}k_1^{k_1}\sqrt{k_2}k_2^{k_2}}{\sqrt{|\bs{k}|_1}|\bs{k}|_1^{|\bs{k}|_1}}=
  \fr{|\bs{k}|_1}\log\frac{\sqrt{k_1k_2}}{\sqrt{|\bs{k}|_1}}+\log\Big(\frac{k_1}{|\bs{k}|_1}\Big)^{\frac{k_1}{|\bs{k}|_1}}
  +\log\Big(\frac{k_2}{|\bs{k}|_1}\Big)^{\frac{k_2}{|\bs{k}|_1}}, 
\end{equation*}
and it is easy to check that $\lim_{|\bs{k}|\ra+\infty}\fr{|\bs{k}|_1}\log\frac{\sqrt{k_1k_2}}{\sqrt{|\bs{k}|_1}}=0$, so that
\begin{equation*}     	\mu_{c;1}(\bs{\y})=\lim_{\substack{|\bs{k}|_1\ra+\infty\\\substack{\bs{k}/|\bs{k}|_1\ra\bs{\y}\\k_1,k_2>0}}}
  \Big[\log\Big(\frac{k_1}{|\bs{k}|_1}\Big)^{\frac{k_1}{|\bs{k}|_1}}+\log\Big(\frac{k_2}{|\bs{k}|_1}\Big)^{\frac{k_2}{|\bs{k}|_1}}+\log g!(|\bs{k}|_1)^{\fr{|\bs{k}|_1}}\Big]=\ls\bs{\y},
  \log\bs{\y}\rs+\mu_c,
\end{equation*}
with the convention $\y_i\log\y_i=0$ if $\y_i=0$ since $x\log x\to 0$ as $x\ra 0$. Finally, points $\bs{k}\in\NNZ^2$ with $k_i=0$
for some $i=1,2$ contribute to the limit only if $\bs{\y}=\bs{e}_i$ for some $i=1,2$. For such points $\bs{k}\in\NNZ^2$, we have
$k_1!k_2!=|\bs{k}|_1!$, and so the first term in the right hand side of~\eqref{BlindTail} vanishes, which agrees with the fact
that $\ls\bs{\y},\log\bs{\y}\rs=0$ if $\bs{\y}=\bs{e}_i$, $i=1,2$. This is to be expected, since in the directions
$\bs{\y}=\bs{e}_i$ with $i=1,2$ in the phase space we have only one of the two species of particles, which when on their own
perform the underlying $1$-species ZRP with critical chemical potential $\hat{\mu}_c=\log\hat{\phih}_c$. This completes the proof
that $\bs{g}$ has regular tails. \qed

\paragraph{Acknowledgement} All authors thank the Leverhulme Trust for its support via grant RPG-2013-261. JZ gratefully
acknowledges funding by the EPSRC through project EP/K027743/1 and a Royal Society Wolfson Research Merit Award.  ND gratefully
acknowledges funding by the EPSRC through project EP/M028607/1. We also thank the anonymous reviewers for their very careful
reading and numerous comments which improved the manuscript.

\def\cprime{$'$} \def\cprime{$'$} \def\cprime{$'$}
  \def\polhk#1{\setbox0=\hbox{#1}{\ooalign{\hidewidth
  \lower1.5ex\hbox{`}\hidewidth\crcr\unhbox0}}} \def\cprime{$'$}
  \def\cprime{$'$}

\end{document}